\documentstyle[12pt,epsf,epsfig]{article}
\def\be{\begin{equation}}
\def\ee{\end{equation}}
\def\ba{\begin{eqnarray}}
\def\ea{\end{eqnarray}}

\def\bq{\begin{quote}}
\def\eq{\end{quote}}

 at 10truept

\newcommand{\beq}{\begin{equation}}
\newcommand{\eeq}{\end{equation}}
\newcommand{\beqa}{\begin{eqnarray}}
\newcommand{\eeqa}{\end{eqnarray}}

\def\ltap{\ \raise.3ex\hbox{$<$\kern-.75em\lower1ex\hbox{$\sim$}}\ }
\def\gtap{\ \raise.3ex\hbox{$>$\kern-.75em\lower1ex\hbox{$\sim$}}\ }
\def\gl{\ \raise.5ex\hbox{$>$}\kern-.8em\lower.5ex\hbox{$<$}\ }
\def\roughly#1{\raise.3ex\hbox{$#1$\kern-.75em\lower1ex\hbox{$\sim$}}}

\begin{document}

\begin{titlepage}
\vfill
\begin{flushright}
\end{flushright}

\vfill
\begin{center}
\baselineskip=16pt
{\Large\bf  AdS$_4$/CFT$_3$+Gravity for Accelerating Conical Singularities  }
\vskip 1.0cm
{\large {\sl }}
\vskip 10.mm
{\bf Mohamed M. Anber\footnote{\tt
manber@physics.umass.edu} } \\
\vskip 1cm
{

       Department of Physics\\
       University of Massachusetts\\
       Amherst, MA 01003\\
}
\vspace{6pt}
\end{center}
\vskip 0.5in
\par
\begin{center}
{\bf ABSTRACT}
\end{center}
We study the quantum-mechanical corrections to two  point particles accelerated by a strut in a $2+1$ D flat background. Since the particles are accelerating, we use finite temperature techniques to compute the Green's function of a conformally coupled scalar applying transparent and Dirichlet boundary conditions at the location of the strut. We find that the induced energy-momentum tensor diverges at the position of the strut unless we impose transparent boundary conditions. Further, we use the regular form of the induced energy-momentum tensor to calculate the gravitational backreaction on the original space. The resulting metric is a constant $\phi$ section of the 4-dimensional C-metric, and it describes two black holes  corrected by weakly coupled CFT and accelerating in asymptotically flat spacetime. Interestingly enough, the same form of the metric was obtained before in 0803.2242 by cutting the AdS C-metric with angular dependent critical 2-brane. According to AdS/CFT+gravity conjecture, the latter should describe strongly coupled CFT  black holes accelerating on the brane. The presence of the CFT  at finite temperature gives us a unique opportunity to study the AdS/CFT+gravity conjecture at finite temperatures.  We calculate various thermodynamic parameters to shed light on the nature of the strongly coupled CFT. This is the first use of the duality in a system containing accelerating particles on the brane.

\begin{quote}

\vfill
\vskip 2.mm
\end{quote}
\end{titlepage}

\section{Introduction}

Since the original construction of the Randall-Sundrum model \cite{Randall:1999ee}, many authors aimed to find black hole solutions localized on a brane (see e.g. \cite{Chamblin:1999by}-\cite{Gregory:2008br}, and \cite{Gregory:2008rf} for a review). Although this is still an open problem in the case of a 3-brane, Emparan, Horowitz and Myers found a class of brane-localized black holes in the lower dimensional case of a 2-brane embedded in four dimensional Anti-de Sitter space  $AdS_4$ \cite{Emparan:1999wa,Emparan:1999fd}. This was found by appropriately cutting the AdS-C metric \cite{Kinnersley:1970zw,Plebanski:1976gy}, which describes bulk accelerated black holes in $AdS_4$, with a brane. Since the tension of this brane is determined by its acceleration in the $AdS_4$ background, this tension is carefully chosen such that the brane acceleration matches the acceleration of the bulk black hole. Recently, this condition was relaxed by detuning the brane tension from the bulk black hole acceleration to find two new classes of solutions on the 2-brane \cite{Anber:2008qu}. The first class has time-dependent induced metrics and describes an evolving  lump of dark radiation, while the second is a constant $\phi$-section of the original four-dimensional C-metric and  describes accelerated black holes by means of strings or struts.

Moreover, it was conjectured in \cite{Emparan:2002px} that black hole solutions localized on a brane in the $AdS_{D+1}$ braneworld correspond to quantum-corrected black holes in $D$ dimensions. This conjuncture followed naturally from the AdS/CFT correspondence in which classical dynamics in the $AdS_{D+1}$ bulk encodes the quantum dynamics of the dual D-dimensional conformal field theory (CFT). Thus, solving the classical $D+1$ classical equations in the bulk is equivalent to solving Einstein equations $G_{\mu\nu}=8\,\pi\,G_D \left\langle T_{\mu\nu}\right\rangle$ on the brane, where $G_D$ is the D-dimensional Newton's constant and $\left\langle T_{\mu\nu}\right\rangle$ is the energy-momentum tensor of a strongly coupled  CFT with a cutoff in the ultraviolet due to the presence of the brane. This conjecture was put to test in \cite{Emparan:2002px} by comparing the brane-black hole solution found in \cite{Emparan:1999wa} with the one obtained in the dual $2+1$ CFT coupled to gravity. This can be done by starting with a point particle of mass $M$ in $2+1$ dimension which generates a deficit angle $\delta=8\,\pi\,G_3 M$, and computing the Casimir energy-momentum tensor \cite{Souradeep:1992ia}. Then, one can use this energy-momentum tensor to calculate the backreaction on the conical spacetime \cite{Soleng:1993yh}. The agreement between the two sides is remarkable and gives a strong argument in favor of the AdS/CFT duality in the context of braneworld scenario. Indeed, this conjecture gives us a convenient way to learn about strong quantum effects in curved backgrounds \cite{Anderson:2004md,Grisa:2007qv,Pujolas:2008rc}.

According to this conjecture, the accelerated black hole solution found in \cite{Anber:2008qu} was interpreted as quantum corrected black hole(s) accelerating on the brane. In the case of a critical brane, this solution describes a pair of  black holes accelerated by two strings (or a strut) pulling (pushing) them toward infinity. In the CFT picture, the energy per unit length of this string (strut) has both classical and quantum contributions. Indeed, while in pure $2+1$ dimensional gravity point particles do not interact, quantum effects generate a force between the particles. Hence, the tension of the string (strut) takes into consideration both effects.

In the present work, we calculate the quantum-mechanical backreaction on two point particles attached to a strut and accelerating in  $2+1$ D flat background. In our setup, we compute the Green's function of a  conformally coupled scalar. Since we work in an accelerating frame moving with acceleration $A$, it is natural to use a quantum field in equilibrium with a thermal bath at temperature $T=A/2\,\pi$, which is the Hawking-Unruh temperature associated with the Rindler (acceleration) horizon. In calculating the thermal Green's function, we impose two different boundary conditions at the location of the strut, namely, transparent and Dirichlet boundary conditions. We find that the numerical coefficient of the induced energy-momentum tensor in the first case is identical to the case of a point particle at rest in a $2+1$ D background. On the other hand, the energy-momentum tensor in the case of Dirichlet boundary conditions is divergent at the position of the strut. This suggests that unless we impose transparent boundary conditions, the location of the strut is susceptible  to the formation of curvature singularity. Further, we use the resulting energy-momentum tensor calculated in the case of transparent boundary conditions to find the gravitational backreaction on the spacetime. Interestingly enough, we find that the resulting quantum corrected metric has the same form as the one obtained before in \cite{Anber:2008qu} by cutting the $AdS_4$ C-metric with a critical angular dependent brane. Although the former solution describes accelerating black hole dressed with {\it weakly} coupled CFT (WCFTBH), the latter, according to AdS/CFT+gravity conjecture, describes accelerating black hole dressed with {\it strongly} coupled CFT (SCFTBH). Moreover, the presence of the CFT at finite temperature gives us a unique opportunity to study finite temperature effects in strongly coupled system in curved background. Contrary to the case of the static black hole constructed previusly in \cite{Emparan:1999wa}, where it was found that the black hole mass can be as large as $1/4G_3$, the largest mass one can place in $2+1$ D, we find that the maximum mass in the present case is temperature dependent. Studying the behavior of this maximum mass reveals a striking difference between the weakly and strongly coupled theories. Although the mass in the former decreases monotonically with temperature from $1/4G_3$ to zero, we find that in the second case it decreases from  $1/4G_3$ at low temperatures to a minimum value, and then  it increases to $1/4G_3$ again at high temperatures. Further, the black hole horizon circumference diverges in both cases as the mass reaches its maximum value. Beyond this mass, the horizon disappears leaving behind a naked singularity. Actually, it was argued before that quantum effects dresses conical singularities with regular horizon given that these singularities are sufficiently massive. This is known as quantum cosmic censorship \cite{Emparan:2002px}.  Although this still applies in  the case of accelerating  conical singularities, supermassive singularities (exceeding the maximum allowed black hole mass at a given temperature) violate this censorship. 

We start our treatment using a classical background with a vanishing value of the total mass of particles and strut $m_{p}+m_{s}=0$. After computing the quantum-mechanical backreaction on the spacetime, we find that the mass of the strut gets renormalized which in turn violates the above equality. The violation is minimal for small values of the black hole mass, and becomes stronger for larger values. 

Our paper is organized as follows. In section 2 we discuss the background geometry used in the setup. Then, in sections 3 and 4 we calculate the thermal Green's function and the induced energy-momentum tensor for both transparent and Dirichlet boundary conditions. The gravitational backreaction due to the weakly coupled CFT is computed in section 5, while in section 6 we show that the same form of the metric found in section 5 can be obtained by cutting the AdS C-metric in 4-D with an angular dependent critical brane \cite{Anber:2008qu}. The latter describes accelerated black hole dressed with strongly coupled CFT. In section 7 we compute various thermodynamic quantities  of the black hole and comment on the AdS/CFT+gravity interpretation of the strongly coupled solution, and finally we conclude in section 8.

\section{Background geometry}

We start by considering the following  metric which describes an accelerating frame moving in a  $2+1$ D flat background with acceleration $A$
\begin{equation}\label{accelerating conical sing}
ds^2=\frac{1}{A^2(w-v)^2}\left[-(v^2-1)dt^2+\frac{dv^2}{v^2-1}+\frac{dw^2}{1-w^2} \right]\,.
\end{equation}
In order for the metric to have Lorentz signature, we restrict $w$ to lie in the range $-1 \le w \le 1$. Also, we restrict $v$ to satisfy $v<w$ since the conformal factor $(v-w)^2$ implies that the points $v=w$ are infinitely far from those $v \ne w$. Moreover, there is a Rindler horizon located at $v=-1$ which is a manifestation of the fact that the metric (\ref{accelerating conical sing}) is written in accelerating coordinates.  This can be shown using the following set of transformations
\begin{eqnarray}
\nonumber
X&=&\frac{\sqrt{v^2-1}}{A(w-v)}\cosh t\,,\\
\nonumber
Y&=&\frac{\sqrt{v^2-1}}{A(w-v)}\sinh t\,,\\
Z&=&\frac{\sqrt{1-w^2}}{A(w-v)}\,,
\label{transformations to flat metric}
\end{eqnarray}
which brings (\ref{accelerating conical sing}) to the flat metric $ds^2=-dY^2+dX^2+dZ^2$. In turn, this restricts $v$ to lie in the range $v \le -1$, and observers in this system can reach asymptotic infinity at $v=w=-1$.

Since gravity is not dynamical in $2+1$ D, the presence of a point mass does not alter the geometry in (\ref{accelerating conical sing}). Instead, it affects the global topology of the spacetime. This can be achieved by introducing a deficit parameter $\delta$ and demanding that $w$ lies in the new range $-1 \le w \le 1-\delta$. The coordinate $v$ is, roughly speaking, the inverse radial direction measured from the location of the particle, and hence those values of $\delta $ different from zero reflect the presence of a point mass located at $v=-\infty$. Since this point mass is accelerating, the acceleration has to be provided  by a co-dimensional $1$ object, a string or strut
\footnote{Although  a co-dimensional $1$ object is formally a brane, we continue to call it a string (strut) in our setup. Strings correspond to positive tension objects, while struts correspond to negative ones. Strings in $2+1$ D were also studied in \cite{Deser:1989cf,Grignani:1989ip}.}
, attached to it. From now on we choose to use a strut located at $w=1-\delta$ which extends from the location of the point mass to the Rindler horizon $v=-1$. This choice renders the system well-behaved at asymptotic infinity. 

\begin{figure}[ht]
\leftline{
\includegraphics[width=0.55\textwidth]{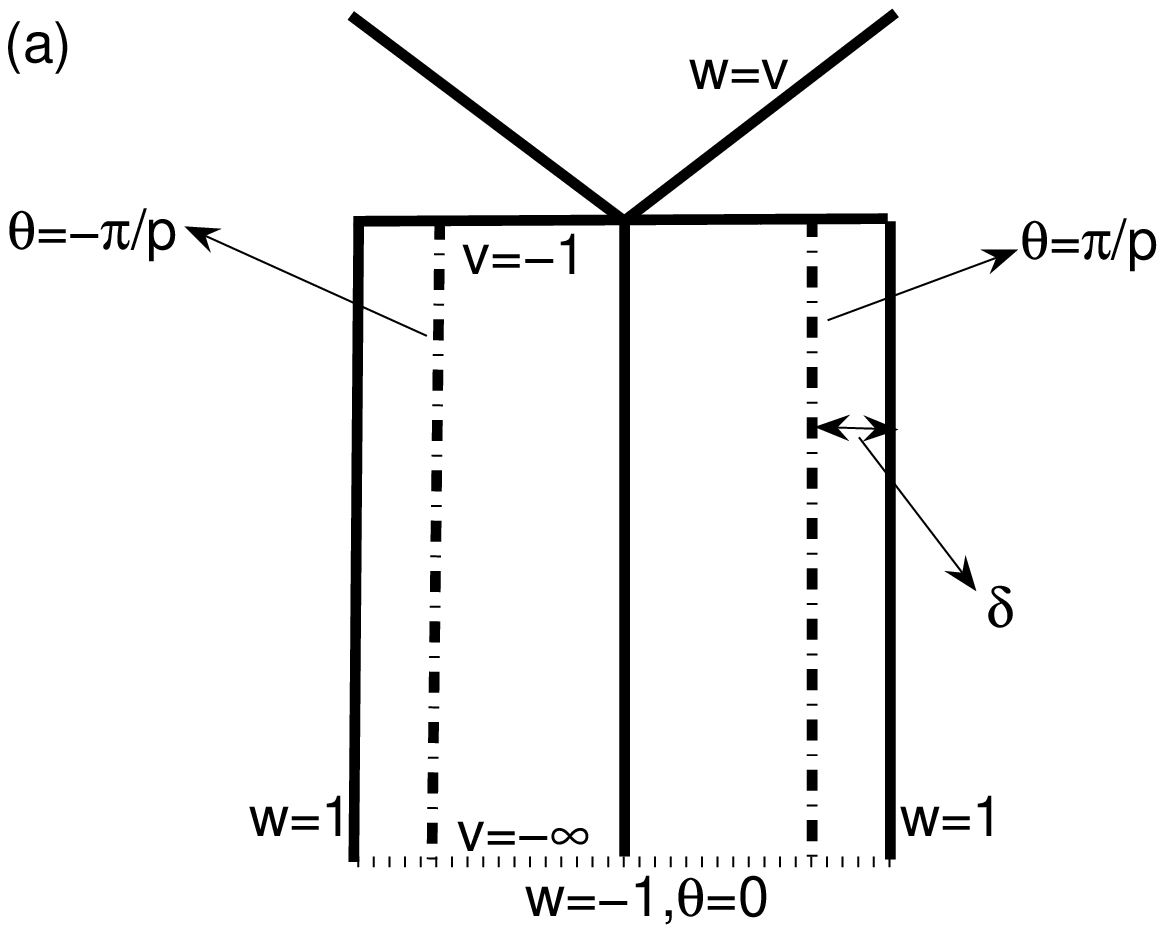}
\includegraphics[width=0.45\textwidth]{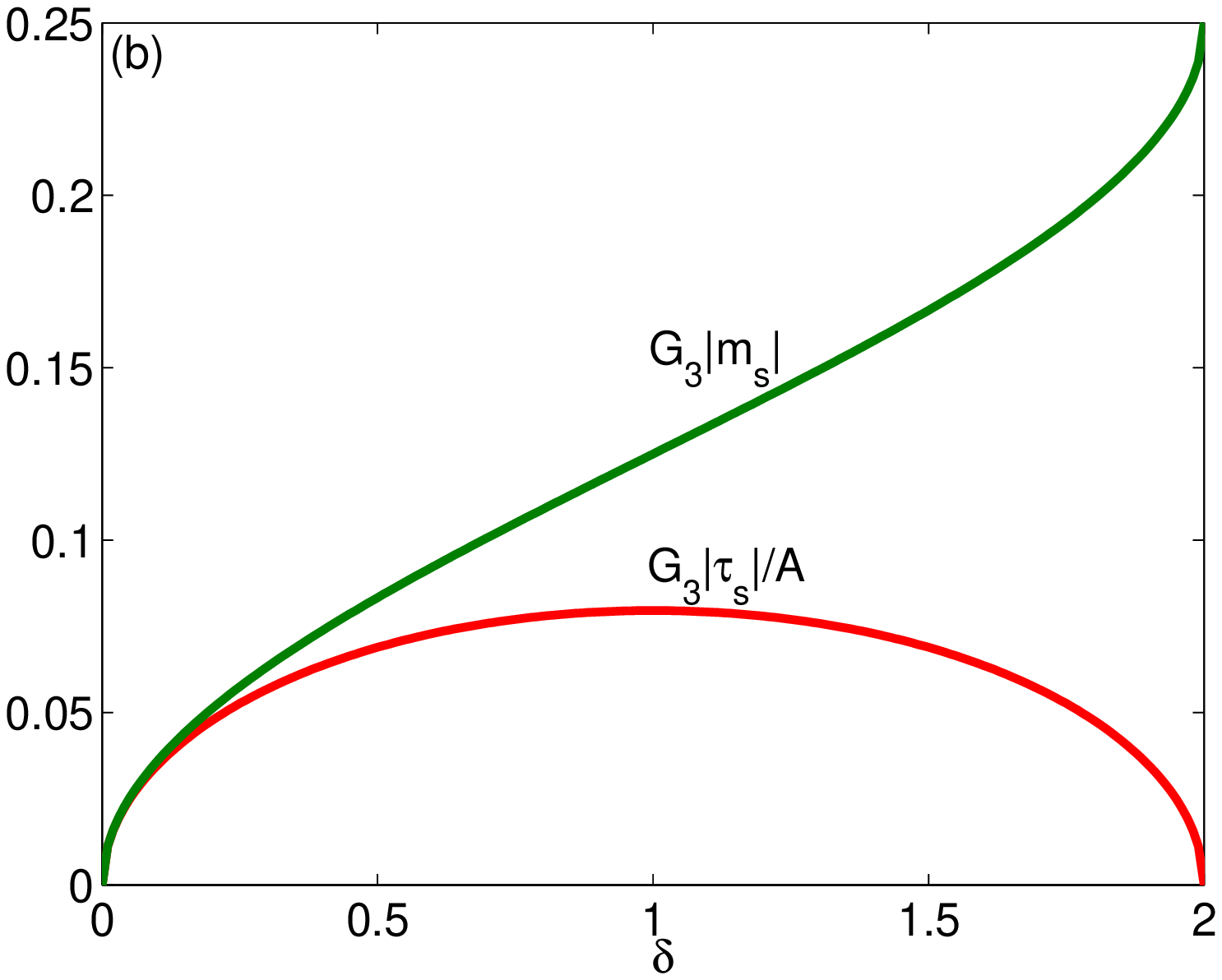}
}
\caption{
(a) A sketch of the background geometry in (\ref{accelerating conical sing}). The strut is located at $w=1-\delta$ which corresponds to $\theta=\pi/p$. The space is $Z_2$ symmetric across $\theta=0$. Notice also that the edges $\theta=\pi/p$ and $\theta=-\pi/p$ are identified.
(b) The absolute value of the mass $|m_s|$ and tension of the strut as  functions of the deficit parameter $\delta$. The tension is a non monotonic function which peaks at $\delta=1$, while the strut mass is monotonic in $\delta$.  
}
\label{space and masses}
\end{figure}

Further, using the toroidal coordinates $v=-\cosh \eta$ and $w=-\cos\theta$, and the Euclidean time $it=\phi$, one can write the metric (\ref{accelerating conical sing}) in the form
\footnote{The coordinates $\theta$, $\eta$ and $\phi$ are known in the mathematical physics literature as toroidal or ring coordinates. See e.g. \cite{bateman,morse}. }
\begin{equation}\label{accelerating mass in ring coordinates}
ds^2=\frac{1}{A^2(\cosh\eta-\cos\theta)^2}\left[ \sinh^2\eta\,d\phi^2+d\eta^2+d\theta^2  \right]\,,
\end{equation} 
where $\eta \ge 0$ and the range $-1 \le w \le 1-\delta$ maps to $0\le \theta \le\pi/p$. The parameter $p$ is defined via $p=\pi/\cos^{-1}(-1+\delta)$ where $p \ge 1$, and $p=1$ corresponds to having empty space. Moreover, we can use the transformations $\cosh \eta=1/Ar$ and $\phi=A\tilde \phi$ that bring the metric (\ref{accelerating mass in ring coordinates}) to the useful form
\begin{equation}
ds^2=\frac{1}{\left(1-Ar\cos\theta\right)^2}\left[\left(1-A^2r^2\right)d\,\tilde \phi^2+\frac{dr^2}{1-A^2r^2}+r^2d\theta^2  \right]\,,
\end{equation}
which is conformal to De Sitter space.
In the limit $A\rightarrow 0$ we find $ds^2=d\tilde \phi^2+dr^2+r^2d \theta^2$, and for empty space $\theta$ lies in the range $-\pi \le \theta \le \pi$. Hence, in the presence of the  strut we restrict the angle $\theta$ in the range $-\pi/p\le \theta \le\pi/p$, which can be covered by gluing two copies of the region $-1 \le w \le 1-\delta$ as sketched in figure(\ref{space and masses}). The mass of the point particle attached to the strut is determined by reminding that in $2+1$ D the mass is given by $2\pi-\Delta\theta=8\pi G_3 m_p $. Hence, we find 
\begin{equation}\label{the mass of the point particle}
m_p=\frac{1}{4G_3}\left[1-\frac{\cos^{-1}\left(-1+\delta\right)}{\pi} \right]\,.
\end{equation}

 On the other hand, one can obtain an expression for the energy per unit length, or the force, $\tau_s$ of the strut. To this end, we write the Israel junction condition for the metric (\ref{accelerating conical sing})
\begin{equation}
\Delta{\cal K}_{\mu\nu}-\ell_{\mu\nu}\Delta{\cal K}=-8\,\pi\,G_3 h_{\mu\nu}\tau_{s}\,,
\end{equation}
where $\Delta{\cal K}_{\mu\nu}={\cal K}_{\mu\nu}^{+}-{\cal K}_{\mu\nu}^{-}$ and $\ell_{\mu\nu}$  are respectively  the jump in the extrinsic curvature and the induced metric at the position of the strut. Hence, we obtain after straight forward calculations noticing the $Z_2$ symmetry across the strut
\begin{equation}\label{classical tension}
\tau_{s}=-\frac{A}{4\,\pi\,G_3}\sqrt{2\delta-\delta^2}\,.
\end{equation}
This equation restricts the range of $\delta$ to be $0 \le \delta \le 2$. In figure (\ref{space and masses}b) we see that there are two values of the deficit parameter $\delta$ that correspond to the same tension. We also find that the minimum tension is $\tau_s=-A/4\,\pi\,G_3$ which occurs at $\delta=1$.

 The mass of the strut is given by  $m_{s}=\tau_{s} \ell_{s}$  where $\ell_{s}$ is the proper length of the strut. The latter can be calculated from (\ref{accelerating conical sing}) 
\begin{equation}
\ell=\frac{1}{A}\int_{-\infty}^{-1}\frac{dv}{(1-\delta-v)\sqrt{v^2-1}}=\frac{\pi-\cos^{-1}(-1+\delta)}{A\sqrt{-\delta^2+2\delta}}\,,
\end{equation}
and hence, we obtain
\begin{equation}
m_s=-\frac{1}{4G_3}\left[1-\frac{\cos^{-1}\left(-1+\delta\right)}{\pi} \right]\,.
\end{equation}
Therefore, we find that the combined mass of the point particle and the strut adds up to zero.  
We also provide another proof of this result in appendix A by working in the $XYZ$ coordinates given in (\ref{transformations to flat metric}). This result is particular to the way we chosed to cut our space and may change upon using more general configurations.

The above construction covers only the half space $X>0$. The other half describes a particle located at $v=\infty$ and accelerating in the opposite direction with acceleration horizon located at $v=1$. This particle is attached to the strut that stretches to the first particle through the horizons.

Finally,  the identification of the Euclidean time coordinate $\phi$ in (\ref{accelerating mass in ring coordinates}) with period $2\,\pi$ gives rise to background temperature $T=A/2\,\pi$. This is Hawking-Unruh temperature (background temperature) as measured by a detector placed in the accelerating background (\ref{accelerating conical sing}).

\section{Scalar field quantization}

We consider a massless scalar field $\Phi$ coupled to a three dimensional background. The corresponding Lagrangian density reads 
\begin{equation}\label{scalar field quantization}
{\cal L}=-\frac{1}{2}\left(\nabla \Phi\right)^2-\frac{\xi}{2} R(x)\,\Phi^2\,,
\end{equation}
where $\xi$ is a numerical factor and $R$ is the Ricci scalar. The values $\xi=0$ and $\xi=1/8$ correspond to minimally and conformally coupled scalars respectively.
Since the Euclidean metric (\ref{accelerating mass in ring coordinates}) is associated to a  background temperature $T=A/2\,\pi$, it is natural to quantize the field $\Phi$ incorporating the non-zero temperature techniques. This can be achieved by going to the Euclidean time \cite{Michel Le Bellac} as we show below.   

The Lagrangian (\ref{scalar field quantization}) leads to the following Euclidean action
\begin{equation}
S_{E}(\beta)=-\frac{1}{2}\int_{0}^{2\,\pi}d\phi\int d^2 x \sqrt{g_{E}}\left[g_{E}^{\mu\nu}\nabla_{\mu}\Phi\nabla_{\nu}\Phi+\xi R(x)\,\Phi^2 \right]\,,
\end{equation}
where $g_{E}$ is the Euclidean metric (\ref{accelerating mass in ring coordinates}). One defines a generating functional $Z(j)$ through
\begin{equation}\label{generating functional}
Z(j)=\int {\cal D}\Phi \mbox{Exp}\left(-S_{E}(\beta)+\int_{0}^{2\,\pi}d\phi \int d^2\, x\sqrt{g_E}j(x)\Phi(x)\right)\,,
\end{equation}  
with $j(x)$ being the external current. The thermal average of the time-ordered product of the field  operators (the thermal propagator) is given by
\begin{equation}
G_F(x,x')=\frac{1}{Z}\frac{\delta^2 Z(j)}{\delta j(x)\delta j(x')}|_{j=0}=U(\phi-\phi')\left\langle  \Phi(x)\Phi(x')\right\rangle+U(\phi'-\phi)\left\langle  \Phi(x')\Phi(x)\right\rangle\,,
\end{equation}
where $U$ is the Heaviside step function, and  the brackets $\langle\,\,\,\rangle$ denote the thermal average. Finally, using the generating functional (\ref{generating functional}) one can show that the propagator $G_F(x,x')$ satisfies the equation 
\begin{equation}\label{Green's function equation}
\left[g_{E}^{\mu\nu}\nabla_{\mu}\nabla_{\nu}-\xi R(x) \right]G_{F}(x,x') =-\frac{1}{\sqrt{g_{E}(x)}}\delta(\phi-\phi')\delta^{2}(x-x')\,.
\end{equation}

Now, we are ready to find the normal-mode sum for the propagator $G_{F}(x,x')$ in the background (\ref{accelerating mass in ring coordinates})
\footnote{For scalar field quantization on the BTZ black hole background \cite{Banados:1992wn} see \cite{Steif:1993zv,Lifschytz:1993eb,Shiraishi:1993hf}. }
. Since this background is flat, the Ricci scalar $R(x)$ vanishes everywhere except at the location of the strut where it is given by $R(x)=8A^2(w_0-v)(1-w_0^2)\delta(w-w_0)$ with $w_0=1-\delta$. Moreover, one can show that eq. (\ref{Green's function equation}) is separable upon using the ansatz
\begin{equation}\label{original ansatz}
G_{F}(x,x')=\sqrt{(w-v)(w'-v')}\,{\cal G}_F(x,x')\,,
\end{equation}
where we have factored out the conformal factor in (\ref{accelerating conical sing}). Plugging the above ansatz in eq. (\ref{Green's function equation}) we obtain in the case of a conformal theory, $\xi=1/8$, a Dirac-delta coefficient that cancels out the contribution from $R(x)$. Hence,  in terms of $\phi\,,\theta\,,\eta$ coordinates given in (\ref{accelerating mass in ring coordinates}) eq.(\ref{Green's function equation}) reads 
\begin{equation}\label{Green's function in theta phi eta}
\left[\frac{\partial}{\sinh \eta\,\partial \eta}\left(\sinh\eta\,\frac{\partial}{\partial\eta} \right)+\frac{\partial^2}{\partial\theta^2}+\frac{1}{\sinh^2\eta^2}\frac{\partial^2}{\partial\phi^2}+\frac{1}{4}\right]{\cal G}_F=-A\delta(\phi-\phi')\delta(\theta-\theta')\delta(\eta-\eta')\,.
\end{equation} 
In the following, we consider different boundary conditions that one can impose on the behavior of the two point function $G_F(x,x')$ at the location of the strut.

\subsection*{Transparent boundary conditions}

In this case, we require the propagator to be continuous across the strut, i.e. $G_F(\theta=\pi/p)=G_F(\theta=-\pi/p)$. Further, by integrating eq. (\ref{Green's function in theta phi eta}) over an interval containing the strut we obtain $\partial{\cal G}_F(\theta=\pi/p)/\partial\theta=\partial{\cal G}_F(\theta=-\pi/p)/\partial\theta$. This means that the strut is actually transparent to the quantum fluctuations of the conformal theory.
Therefore, we find that the normal modes are given by the expansion
\begin{equation}
{\cal G}_F(x,x')=
\frac{p}{4\pi^2}\sum_{n=0}^{\infty}\sum_{m=0}^{\infty}\epsilon_{n}\epsilon_{m}\cos pn(\theta-\theta')\cos m(\phi-\phi')\hat G_F(\eta;n;m)\,,\\
\label{main ansatz}
\end{equation}   
where  $\epsilon_n=2-\delta_{n,0}$. Substituting the above expansion in eq. (\ref{Green's function in theta phi eta}) we obtain
\begin{equation}\label{Legendre equation}
(1-v^2)\frac{\partial^2 \hat G_F }{\partial v^2}-2v\frac{\partial \hat G_F}{\partial v}+\left[ p^2\,n^2-\frac{1}{4}-\frac{m^2}{1-v^2}\right]\hat G_F=A\delta(v-v')\,,
\end{equation}
where $v=-\cosh\eta$. This is associated Legendre equation with well-behaved solution for the range $-v \ge 1$ given by
\begin{equation}
\hat G_F(v,v')=A_{mn}P^{-m}_{pn-1/2}(-v_<)Q^{-m}_{pn-1/2}(-v_>)\,,
\end{equation}
where $-v_<=\mbox{min}\{-v,-v'\}$ and $-v_>=\mbox{max}\{-v,-v'\}$, and $P^{-m}_{pn-1/2}$ and $Q^{-m}_{pn-1/2}$ are associated Legendre functions of the first and second kind respectively
\footnote{For a summary of some properties and relations of associated Legendre functions we use below, see appendix B.}
 \cite{bateman,Stegun,Snow}. To determine the arbitrary constants $A_{mn}$, we integrate eq. (\ref{Legendre equation}) over an interval containing $v'$ and use the Wronskian relation (\ref{Wronskin}) to find $A_{mn}=A(-1)^m\,\Gamma(1/2+m+pn)/\Gamma(1/2-m+pn)$. Further, using the addition theorem (\ref{addition theorem}) we obtain
\begin{eqnarray}
\nonumber
G_{F}(x,x')&=&\frac{p\,A}{4\pi^2}\sqrt{\left(\cosh\eta-\cos\theta\right)\left(\cosh\eta'-\cos\theta' \right)}\sum_{n=0}^{\infty}\epsilon_{n}\cos pn(\theta-\theta')Q_{pn-1/2}(\cosh\gamma)\,,
\label{main series}
\end{eqnarray}
where $\cosh\gamma=\cosh\eta\cosh\eta'-\sinh\eta\sinh\eta'\cos(\phi-\phi')$. 

For the case of empty space, $p=1$, we can use the identity (\ref{identity for Legendre functions}) to find closed-form expression of the propagator
\begin{eqnarray}
\nonumber
G^{0}_F(x,x')&=&\frac{A}{4\sqrt{2}\pi}\frac{\sqrt{\left(\cosh\eta-\cos\theta \right)\left(\cosh\eta'-\cos\theta' \right)}}{\sqrt{\cosh\eta\cosh\eta'-\sinh\eta\sinh\eta'\cos(\phi-\phi')-\cos(\theta-\theta')}}\,.\\
\end{eqnarray}
One can see that $4\pi G^{0}_F(x,x')=1/\sqrt{(X-X')^2+(Y-Y')^2+(Z-Z')^2}$ is the inverse distance between the space points $x$ and $x'$ given in the Euclideanized version of (\ref{transformations to flat metric}). 

The two point function  $G_{F}(x,x')$ is ultraviolet divergent in the coincidence limit $x \rightarrow x'$, and one needs to regularize it before proceeding to real calculations. Generally, this can be achieved using Schwinger-DeWitt expansion of the propagator in powers of the geodesic distance between $x$ and $x'$, and then subtracting the divergent terms which result upon taking the coincidence limit \cite{Birrell:1982ix}. Fortunately, one does not need this tedious procedure here thanks to the absence of trace anomalies in odd dimensions. Simply, the renormalization can be performed by subtracting out the empty-space contribution $G^{0}_F(x,x')$. Therefore after using the integral representation of $Q_{pn-1/2}(z)$ given in (\ref{integral representation of Q}), we find
\begin{eqnarray}
\nonumber
G^{R}_{\mbox{\scriptsize TR}}(x,x')&=&G_{F}(x,x')-G^{0}_{F}(x,x')\\
\nonumber
&=&\frac{A}{4\sqrt{2}\pi^2}\sqrt{\left(\cosh\eta-\cos\theta\right)\left(\cosh\eta'-\cos\theta'\right)}\times\\ 
\nonumber
&\times&\int_{\gamma}^{\infty}\frac{du}{\sqrt{\cosh u-\cosh\gamma}}\left[ \frac{p\,\sinh p\,u}{\cosh p\,u-\cos p(\theta-\theta')}-\frac{\sinh u}{\cosh u-\cos (\theta-\theta')} \right]\,.\\
\label{main result}
\end{eqnarray}
%

\subsection*{Dirichlet boundary conditions}
In this case one requires that the field fluctuations vanish at the position of the strut, i.e. $G_F(\pi/p)=G_F(-\pi/p)=0$. These boundary conditions are satisfied by the expansion (\ref{main ansatz}) after replacing $\epsilon_{n}\cos p\,n(\theta-\theta')\rightarrow 2\cos p(n+1/2)\theta\cos p(n+1/2)\theta'$. Hence, upon going through the same procedure above, we find
\begin{eqnarray}
\nonumber
G^{R}_{\mbox{\scriptsize D}}(x,x')
&=&\frac{A}{8\sqrt{2}\pi^2}\sqrt{\left(\cosh\eta-\cos\theta\right)\left(\cosh\eta'-\cos\theta'\right)}\times\\
\nonumber
&\times&\int_{\gamma}^{\infty}\frac{du}{\sqrt{\cosh u-\cosh\gamma}}\left[ \frac{p\,\cos p\,\theta_{\mbox{\scriptsize av}}\sinh (p\,u/2)}{\sinh^2(p\,u/2)+\sin^2 p\,\theta_{\mbox{\scriptsize av}}}-\frac{\cos\theta_{\mbox{\scriptsize av}}\sinh(u/2)}{\sinh^2(u/2)+\sin^2\theta_{\mbox{\scriptsize av}}}\right.\\ 
\nonumber
&+&\left. \frac{p\,\cos(p\,\Delta\theta/2)\sinh (p\,u/2)}{\sinh^2(p\,u/2)+\sin^2(p\,\Delta\theta/2)}-\frac{\cos(\Delta\theta/2)\sinh(u/2)}{\sinh^2(u/2)+\sin^2(\Delta\theta/2)} \right]\,,
\nonumber
\label{D Green's function}
\end{eqnarray}
where $\Delta\theta=\theta-\theta'$, and $\theta_{\mbox{\scriptsize av}}=(\theta+\theta')/2$.

At this point we are in a position to calculate the renormalized expectation value of the field fluctuations $\left\langle \Phi^2(x) \right\rangle$ as well as the energy-momentum tensor $\left\langle T_{\nu}^{\mu}(x) \right\rangle$.

\section{Quantum stress tensor}

We calculate the expectation value of the field fluctuations as the coincidence limit of the renormalized propagator \cite{Birrell:1982ix}
\begin{equation}
\left\langle \Phi^2(x) \right\rangle=\mbox{lim}_{x\rightarrow x'}G^{R}(x,x')\,,
\end{equation}
which produces upon taking the limit in eq. (\ref{main result}) 
\begin{equation}\label{field expectation value}
G^{R}_{\mbox{\scriptsize TR}}(0)=\frac{A\,I_{1}(p)}{4\sqrt{2}\pi^2}\left(\cosh\eta-\cos\theta\right)\,,
\end{equation} 
where $I_{1}(p)$ is given in appendix C. Interestingly enough, we find that the numerical coefficient in front of the overall factor  $A\left(\cosh\eta-\cos\theta\right)$ is identical to the case of a conical singularity with the same mass sitting at rest in $2+1$ D \cite{Souradeep:1992ia}. Similarly, $G^R(0)$ diverges as $\eta \rightarrow\infty$, the location of the point mass. 
On the other hand, the above limit in the case of Dirichlet boundary conditions gives
\begin{eqnarray}
\nonumber
G^{R}_{\mbox{\scriptsize D}}(0)&=&\frac{A}{8\sqrt{2}\pi^2}\left(\cosh\eta-\cos\theta\right)\int_{0}^{\infty}\frac{du}{\sqrt{\cosh u-1}}\times \\
\nonumber
&\times&\left[\frac{p}{\sinh(p\,u/2)}-\frac{1}{\sinh(u/2)}+ \frac{p\,\cos p\,\theta\sinh (p\,u/2)}{\sinh^2(p\,u/2)+\sin^2 p\,\theta}-\frac{\cos\theta\sinh(u/2)}{\sinh^2(u/2)+\sin^2\theta} \right]\,,\\
\end{eqnarray}
which is divergent at the position of the strut $\theta=\pi/p$. 

\begin{figure}[ht]
\leftline{
\includegraphics[width=0.5\textwidth]{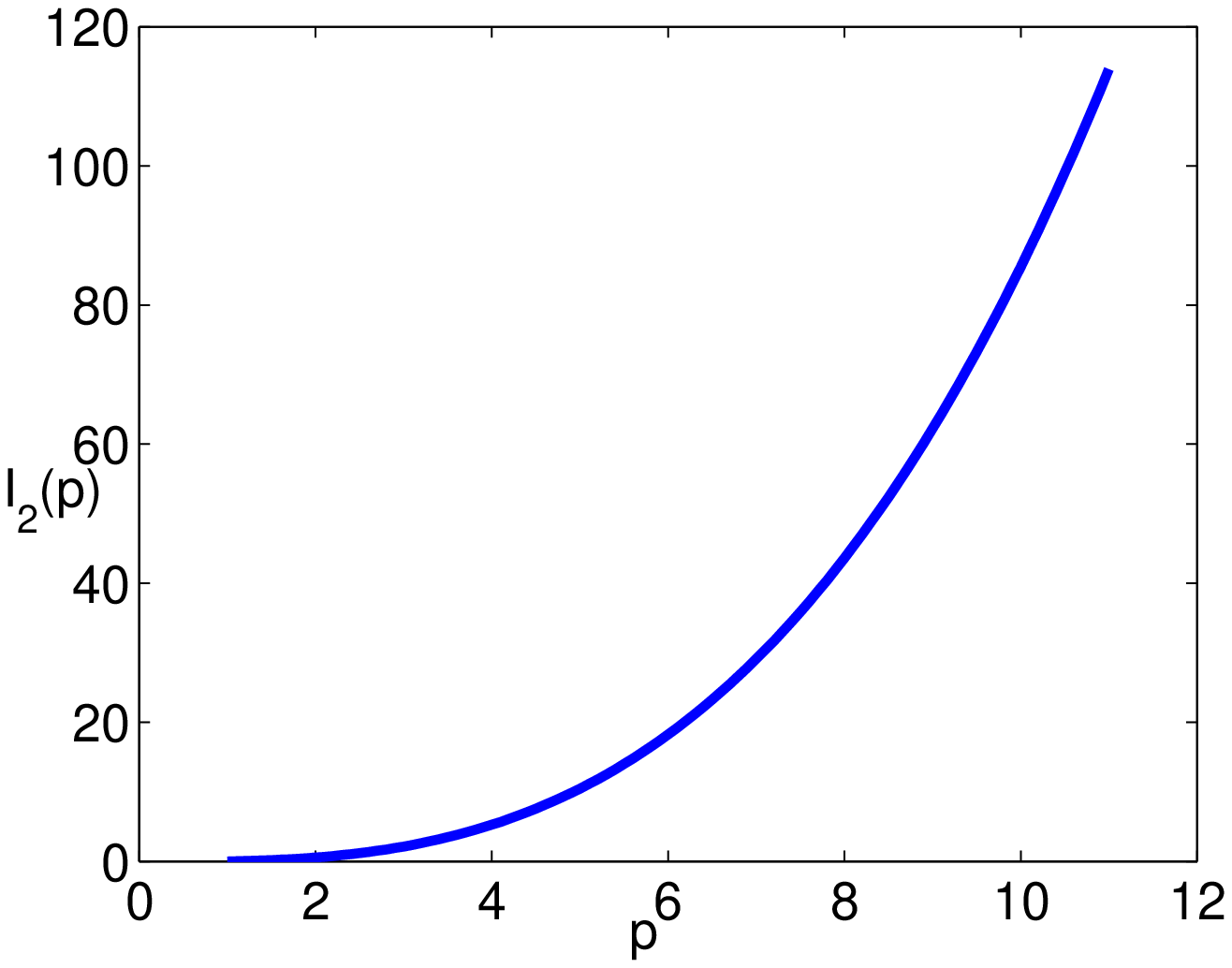}
\includegraphics[width=0.5\textwidth]{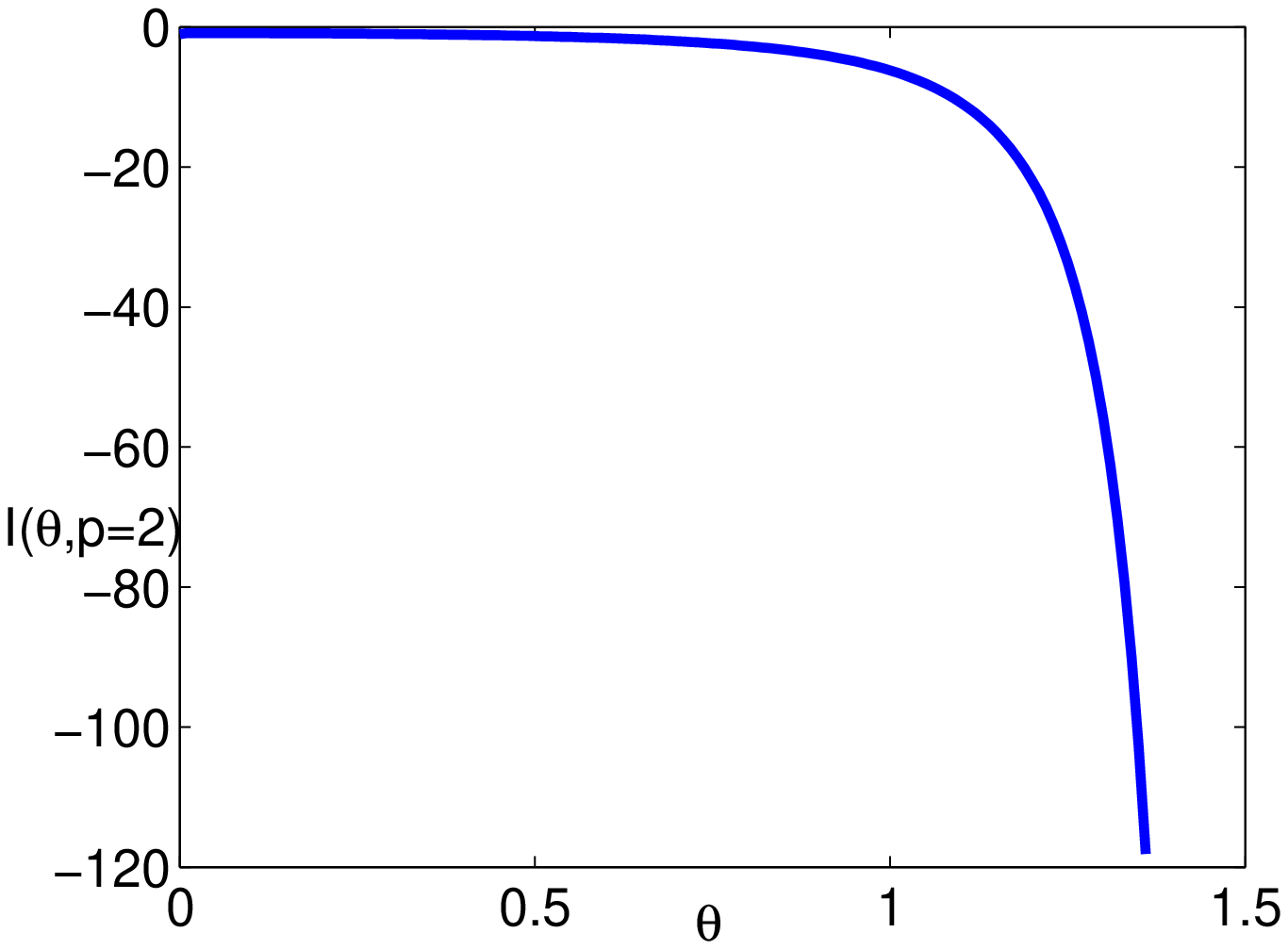}
}
\caption{The numerical coefficients of the energy-momentum tensor in the case of transparent and Dirichlet boundary conditions, $I_2(p)$ and $I(\theta,p)$ respectively. On the left, $I_2(p)$ is plotted as a function of $p$. On the right, $I(\theta,p=2)$ is plotted as a function of $\theta$. This coefficient diverges at the position of the strut $\theta=\pi/p$.
}
\label{plot of I2}
\end{figure}

The expectation value of the energy-momentum tensor for conformally coupled scalar in flat background is given by \cite{Birrell:1982ix}
\begin{equation}\label{coincidence limit}
\left\langle T_{\nu}^{\mu}(x) \right\rangle=\mbox{lim}_{x'\rightarrow x}\left[\frac{3}{4}g^{\mu\lambda}\partial_{\lambda}\partial^{\prime}_{\nu}-\frac{1}{4}g^{\mu}_{\nu}g^{\lambda\beta}\partial_{\beta}\partial^{\prime}_{\lambda}-\frac{1}{4}g^{\mu\lambda}\nabla_{\lambda}\partial_{\nu} \right]G^{R}(x,x')\,.
\end{equation}
Using the coincidence limit of the various derivatives calculated in appendix C we obtain
\begin{equation}\label{the final answer of the energy-momentum tensor}
\left\langle T_{\nu\,{\mbox{\scriptsize TR}}}^{\mu}(x) \right\rangle=\frac{A^3 I_{2}(p)}{8\sqrt{2}\pi^2}\left(\cosh\eta-\cos\theta \right)^3\mbox{diag}(1,1,-2)\,,
\end{equation}
where $I_{2}(p)$ is given in (\ref{integrals per}) (see figure (\ref{plot of I2})). Once more, the numerical factor in front of  $A\left(\cosh\eta-\cos\theta\right)$ is equal to that in the case of a conical singularity at rest. This energy-momentum tensor is not of a thermal type $\propto \mbox{diag}(-2,1,1)$ although the background used to derive the Green's function (\ref{main result}) is in a thermal state. This reflects the presence of a large Casimir component as we have  $\left\langle T_{00\,{\mbox{\scriptsize TR}}}(x) \right\rangle<0$, which violates the strong energy condition.

Taking the coincidence limit (\ref{coincidence limit}) for the case of Dirichlet boundary conditions we find
\begin{equation}\label{ energy-momentum tensor D}
\left\langle T_{\nu\,{\mbox{\scriptsize D}}}^{\mu}(x) \right\rangle=\frac{A^3 I(\theta,p)}{16\sqrt{2}\pi^2}\left(\cosh\eta-\cos\theta \right)^3\mbox{diag}(1,1,-2)\,,
\end{equation}
where $I(\theta,p)$ is given by (\ref{integral D}). The factor $I(\theta,p)$ is divergent at  the location of the strut $\theta=\pi/p$ which is a general property for the energy-momentum tensor with Dirichlet boundary conditions near a curved surface \cite{Birrell:1982ix,Kennedy:1979ar}. Since for a conformal energy-momentum tensor one has $T_{\mu\nu}T^{\mu\nu}\propto R_{\mu\nu}R^{\mu\nu} $, we find that this divergent behavior indicates that the location of the strut is susceptible to the formation of a curvature singularity unless we impose transparent boundary conditions.

In the following, we use $\left\langle T_{\nu{\mbox{\scriptsize }}}^{\mu}(x) \right\rangle$ attempting to find the gravitational backreaction using the semi-classical approximation. We limit our analysis to the case of transparent boundary conditions.

\section{Gravitational backreaction}

In this section, we use the regularized energy-momentum tensor derived in section 4 in the case of transparent boundary conditions to solve the semi-classical Einstein equations 
\begin{equation}\label{Einstein equations }
G_{\mu\nu}=8\,\pi\,G_3 \left\langle T_{\mu\nu}(x) \right\rangle\,.
\end{equation}
%
\begin{figure}[ht]
\centerline{
\includegraphics[width=0.5\textwidth]{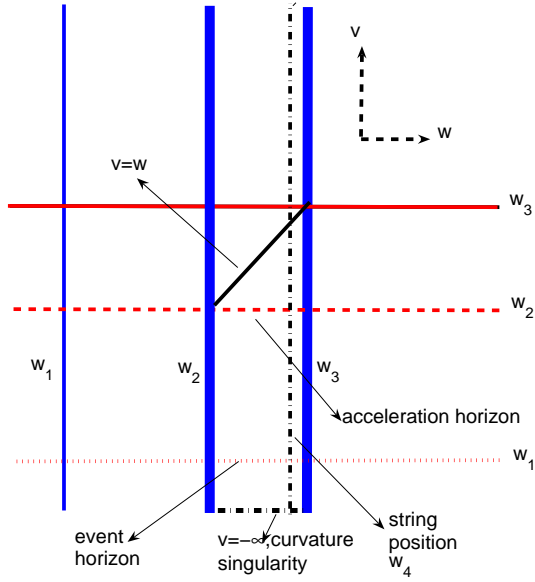}
}
\caption{A sketch of the parameter space of the black hole solution (\ref{black hole on brane}) with $\mu <1/3\sqrt{3}$. The values of $w$ lie in the range $w_2\le w \le w_4$. There is a black hole horizon at $v=w_1$, and acceleration horizon at $v=w_2$.
}
\label{black hole with horizon}
\end{figure}
Since $\left\langle T_{\nu}^{\mu}(x) \right\rangle$ has the form (\ref{the final answer of the energy-momentum tensor}), we find that the most general metric that respects this symmetric form is given by
\begin{equation}\label{semi-classical ansatz}
ds^2=\frac{1}{A^2(\cosh\eta-\cos\theta)^2}\left[-f(\eta)\sinh^2\eta\, dt^2+\frac{d\eta^2}{f(\eta)}+\frac{d\theta^2}{g(\theta)} \right]\,,
\end{equation}
where we have switched back to the Lorentzian time. Substituting the ansatz (\ref{semi-classical ansatz}) into (\ref{Einstein equations }), it is not difficult to show that the solution is given by 
\begin{eqnarray}
\nonumber
f(\eta)&=&1-2\mu\,\frac{\cosh^3 \eta}{\sinh^2 \eta}\,,\\
g(\theta)&=&1+2\mu\,\frac{\cos^3\theta}{\sin^2\theta}\,,
\end{eqnarray} 
where $\mu=G_3 AI_{2}(p)/\sqrt{2}\,\pi$. Using the coordinates $v=-\cosh\eta$ and $w=-\cos\theta$ we finally obtain
\begin{equation}\label{black hole on brane}
ds^2=\frac{1}{A^2(w-v)^2}\left[-(-1+v^2+2\mu\,v^3)\,dt^2+\frac{dv^2}{-1+v^2+2\mu\,v^3}+\frac{dw^2}{1-w^2-2\mu\,w^3} \right]\,.
\end{equation} 
This solution is an equatorial section of the four dimensional C-metric \cite{Kinnersley:1970zw}, and it describes quantum  accelerated black holes, with mass given by (\ref{the mass of the point particle}), dressed with weakly coupled CFT in asymptotically flat spacetime. 

The function $1-w^2-2\mu\,w^3$  has three zeros $w_1<w_2<0<w_3$, provided that $\mu <1/3\sqrt{3}$.  To have Lorentz signature, the range of $w$ is restricted in the range $w_2 \le w \le w_4$ where $w_4<w_3$ is the location of the strut. Moreover, $v$ has to satisfy $v<w$ , and hence we restrict $v$ in the range $-\infty<v<w$. Therefore, observers in the accelerating system can reach  asymptotic infinity  at $w=v=w_2$. As in figure (\ref{black hole with horizon}),  there is a curvature singularity at $v=-\infty$ dressed  by a black hole horizon at $w_1$. Also, there is acceleration horizon at $v=w_2$. In  section 7, we will see that as one approaches $\mu \rightarrow 1/3\sqrt{3}$, the position of the strut $w_4$ is pushed closer to the zero $w_2$, and in the same time the event horizon at $w_1$ closes up with the acceleration horizon at $w_2$, until the range of $v$ disappear when $\mu= 1/3\sqrt{3}$.
 If $\mu> 1/3\sqrt{3}$, then $1-w^2-2\mu\,w^3$ has only one real zero. In this case, there is a naked curvature singularity to all observers (see \cite{Farhoosh:1980ek} for the C-metric in $3+1$ D.) The situation here is completely different from a static point particle sitting in $2+1$ D, where it was found that quantum corrections from a CFT dress the singularity  with a regular horizon. This is known as {\em quantum cosmic censorship}. The analysis above shows that  supermassive accelerating conical  singularities in $2+1$ D violate the quantum cosmic censorship. More on this is in section 8.

 The metric (\ref{black hole on brane}) was obtained before  as a brane solution that results by cutting the AdS C-metric with angular dependent critical brane \cite{Anber:2008qu}. In the next section, we review this construction and then we use the AdS/CFT correspondence to shed some light on the strongly coupled CFT.

\section{The AdS construction}

In this section, we review the construction of the solution (\ref{black hole on brane}) using the braneworld scenario \cite{Anber:2008qu}. We start with the AdS C-metric in $4$ D written in the form \cite{Plebanski:1976gy}
\begin{equation}\label{4 D C-metric}
ds^2=\frac{1}{\tilde A^2(x-y)^2}\left[-H(y)d\psi^2+\frac{dy^2}{H(y)}+\frac{dx^2}{G(x)}+G(x)d\Phi^2\right]\,,
\end{equation} 
where the functions $H(y)$ and $G(x)$ are given by
\begin{eqnarray}
\nonumber
H(y)&=&\lambda+y^2+2G_4 M\tilde A y^3\,,\\
G(x)&=&1-x^2-2G_4 M\tilde A x^3\,.
\end{eqnarray}
In the above expression we take $\lambda>-1$, and $G_4$ denotes the four dimensional Newton's constant. This metric describes one or two accelerated black holes with mass $M$ and acceleration $\tilde A$ on Anti-de Sitter (AdS) background with radius $\ell_4=1/\tilde A\sqrt{\lambda+1}$. As explained in section 5, the function $G(x)$ has three roots $x_1<x_2<0<x_3$ provided that $G_4 M\tilde A<1/3\sqrt{3}$. The direction defined by $x_2$ and $x_3$ corresponds to the axis of rotation. To avoid a conical singularity at $x_3$, we take $\Phi$ to have the period $\Delta\Phi_{\mbox{\scriptsize{bulk}}}=4\pi/|G'(x_3)|$. This leaves a conical singularity at $x_2$ which is interpreted in this case as a cosmic string in the bulk that extends from the black hole out to the AdS boundary.

\begin{figure}[ht]
\centerline{
\includegraphics[width=0.5\textwidth]{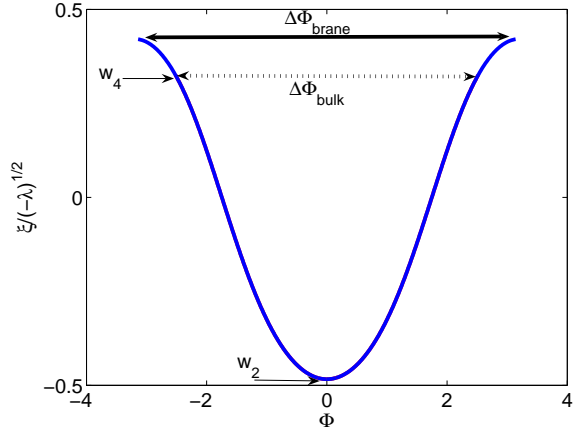}
}
\caption{The numerical solution of eq. (\ref{equation of the brane}) where we take $\lambda=-0.2$ and $G_4M\tilde A=0.15$. The solid and the dotted arrows indicate the period of the angular variable $\Phi$ on the brane and in the bulk, respectively. The value $w_2$ is the largest negative zero of $G(\xi/\sqrt{-\lambda})-\lambda-1$ or equivalently, by means of the transformations (\ref{transformation of induced BH}), $1-w^2-2G_4MAw^3$. The strut is located at $w_4$ where the range of $\Phi$ coincides with the full period of the bulk.  }
\label{period on the brane}
\end{figure}

Next, we look for a purely tensional and critical (asymptotically flat) $Z_2$ symmetric 2-brane embedded  in the above bulk. This brane is described by the surface $x=\xi(\Phi)$
\footnote{The special case $\lambda=0$ was studied previously in \cite{Emparan:1999wa}. This corresponds to cutting the bulk with the brane $x=0$. The brane-induced metric was found to describe a $2+1$ Schwarzschild-like static black hole.}
. The Israel junction conditions on this surface read
\begin{equation}
\Delta{\cal K}_{ab}=-8\pi\,G_4\left[S_{ab}-\frac{1}{2}Sh_{ab}\right]\,,
\end{equation}
where $\Delta {K}_{ab}=K^{+}_{ab}-K^{-}_{ab}$ is the jump in the extrinsic curvature, and $S_{ab}$ is the energy-momentum tensor localized on the brane. For purely tensional and critical brane we have $S_{ab}=h_{ab}/2\pi G_4 \ell_{4}$, where $h_{ab}$ is the surface induced metric. Using Israel junction conditions, we find that $\xi(\Phi)$ obeys the differential equation 
\begin{equation}\label{equation of the brane}
\left(\frac{d \xi}{d \Phi}\right)^2=\frac{G(\xi(\Phi))^3}{1+\lambda}-G(\xi(\Phi))^2\,,
\end{equation}
as well as the auxiliary equation
\begin{equation}
2G(\xi)\frac{d^2\xi}{d\Phi^2}-3G'(\xi)\left(\frac{d \xi}{d \Phi}\right)^2-G^2(\xi)G'(\xi)=0\,.
\end{equation}
Although there is no closed-form solution of eq. (\ref{equation of the brane}), numerical integration shows that $\xi$ is periodic in $\Phi$ with period that is always larger than that of the bulk, i.e. $\Delta \Phi_{\mbox{\scriptsize brane}}>\Delta \Phi_{\mbox{\scriptsize bulk}}$ as shown in figure (\ref{period on the brane}).  This discrepancy between the two periods indicates the existence of a codimensional one object, a strut, on the brane. As was shown in \cite{Anber:2008qu}, the tension of this strut is given by
\begin{equation}\label{tension for strongly coupled CFT}
\tau=-\frac{\tilde A}{4\pi G_3}\sqrt{-1-\lambda+G(\xi(\Delta \Phi_{\mbox{\scriptsize bulk}}/2))}\,.
\end{equation} 

Using $x$ as independent variable in (\ref{4 D C-metric})  we obtain the brane-induced metric
\begin{equation}
ds^2=\frac{1}{\tilde A^2(x-y)^2}\left[-H(y)d\psi^2+\frac{dy^2}{H(y)}+\frac{dx^2}{-1-\lambda+G(x)}\right]\,.
\end{equation}
Finally, one can use the coordinate transformations
\begin{eqnarray}
\nonumber
w&=&\frac{x}{\sqrt{-\lambda}}\,,\qquad v=\frac{y}{\sqrt{-\lambda}}\,,\qquad t=\sqrt{-\lambda}\,\psi\\
A&=&\tilde A\sqrt{-\lambda}\,,\label{transformation of induced BH}
\end{eqnarray} 
to bring the above  metric to the form 
\begin{equation}\label{black hole on brane strong CFT}
ds^2=\frac{1}{A^2(w-v)^2}\left[-(-1+v^2+2G_4 MA\,v^3)\,dt^2+\frac{dv^2}{-1+v^2+2G_4MA\,v^3}+\frac{dw^2}{1-w^2-2G_4MA\,w^3} \right]\,,
\end{equation} 
which is exactly  (\ref{black hole on brane}) with $\mu$ being replaced by $G_4 M A$. However, contrary to (\ref{black hole on brane}) which describes an accelerated black hole dressed with weakly coupled CFT (WCFTBH), the solution (\ref{black hole on brane strong CFT}), according to \cite{Emparan:2002px}, should describe a black hole dressed with strongly coupled CFT (SCFTBH). Finally, the full spacetime  consists of gluing two copies of the region $\xi(\Phi) \le x \le x_3$ along the surface $x=\xi(\Phi)$. In the next section we proceed to shed more light on  WCFTBH and SCFTBH solutions by comparing various thermodynamic properties of the two spaces.

\section{ Black hole thermodynamics and AdS/CFT+gravity interpretation}

In this section, we will elaborate more on comparing the behavior of the two solutions (\ref{black hole on brane}) and (\ref{black hole on brane strong CFT}). To this end, one  needs to calculate the black hole mass of the SCFTBH given in (\ref{black hole on brane strong CFT}). Since the space in (\ref{black hole on brane strong CFT}) is written in accelerating coordinates, one can not define a mass in the usual way. Nevertheless, we may proceed using the coordinate transformations
\begin{eqnarray}
\nonumber
R=-\frac{1}{Av}\,,\qquad T=\frac{t}{A}\,,\qquad d\Theta=\frac{dw}{\sqrt{1-w^2-2G_4MA w^3}}\,,
\end{eqnarray} 
that bring the metric in (\ref{black hole on brane strong CFT}) to the form
\begin{equation}\label{  conformal ads Schwarz}
ds^2=\frac{1}{\left(1+ARw(\Theta)\right)^2}\left[-\left(1-A^2R^2-\frac{2G_4M}{R}\right)dT^2+\frac{dR^2}{1-A^2R^2-\frac{2G_4M}{R}}+R^2d\Theta^2 \right]\,,
\end{equation}
which is conformal to Schwarzschild-De Sitter spacetime. Now, let us assume  the parameters $A$ and $G_4M$ are chosen such that there is an intermediate region where $1<<AR$ and $1<<2G_4M/R$. In this region  the mass of the black hole in  $2+1$ D is given by $m_{\mbox{\scriptsize BH}}=(2\pi-\Delta\Theta)/8\pi G_3$ with
\begin{equation}\label{definition of theta}
\Delta\Theta=2\int_{w_2}^{w_4}\frac{dw}{\sqrt{1-w^2-2G_4MA w^3}}\,,
\end{equation}
where as before $w_2$ and $w_4$ are the largest negative zero of $1-w^2-2G_4A w^3$ and the position of the strut, respectively. This is our definition of the mass and we will continue to use it for all values of  $A$ and $G_4M$. Further, to make a connection to WCFTBH, we define the deficit parameter $\delta$ that corresponds to $m_{\mbox{\scriptsize{BH}}}$ as $\delta=1+\cos(\pi/p)$ where $p=2\pi/\Delta\Theta$. One can also invert eq.(\ref{definition of theta}) to solve for the position of the strut in terms of $\Theta$ in the WCFTBH case after replacing $G_4MA$ with $\mu$.

Moreover, the proper circumference of the event horizon measured by an observer localized on the brane is given by
\begin{equation}
\mbox{C}=\int_{w_2}^{w_4}\frac{dw}{{A(w-w_1)}\sqrt{1-w^2-2G_4MAw^3}}\,.
\end{equation}
This observer would ascribe a black hole entropy of $S_3=\mbox{C}/4G_3$. However, since the black hole horizon extends off the brane, the $4$ D entropy as measured by a bulk observer can be calculated directly from (\ref{4 D C-metric})
\begin{equation}
S_4=\frac{2}{4G_4\tilde A^2}\int_{\Phi=0}^{\Phi=\Delta\Phi_{\mbox{\scriptsize bulk}}}d\Phi\int_{x=\xi(\Phi)}^{x=x_3}\frac{dx }{(x-\sqrt{-\lambda} w_1)^2}\,,
\end{equation}
where we have included a factor of $2$ as we glue two copies of the bulk along the brane. The two expressions $S_3$ and $S_4$ should not agree in general. However, we expect, as in the case of the static $2+1$ D brane-black hole \cite{Emparan:1999wa}, $S_3$ to yield the $S_4$ result for large horizons which occurs as $G_4MA \rightarrow 1/3\sqrt{3}$, as we show below.

One can also use the Euclidean version of (\ref{  conformal ads Schwarz}) to obtain the black hole and  acceleration temperatures by computing the Euclidean time periods required to avoid conical singularities, in the $T-R$ plane, at one of the horizons  
\begin{eqnarray}
\nonumber
T_{\mbox{\scriptsize BH}}&=&\frac{A(w_2-w_1)(w_3-w_1)}{4\pi\, w_1w_2w_3}\,,\\
T_{\mbox{\scriptsize ACC}}&=&\frac{A(w_2-w_1)(w_3-w_2)}{4\pi\, w_1w_2w_3}\,.
\end{eqnarray}
In the limit of small $A$, $T_{\mbox{\scriptsize ACC}}$ can be approximated as
\begin{equation}
  T_{\mbox{\scriptsize ACC}} \approx T\left(1-2G_4MA+{\cal O}\left((G_4MA)^2\right) \right)\,
\end{equation}
 where we have used the background temperature $T=A/2\pi$. This shows that the acceleration temperature is almost a constant and equal to the background temperature  for large range of $G_4MA$. In the same limit we find that the black hole temperature is given by 
\begin{eqnarray}
T_{\mbox{\scriptsize BH}}\approx \frac{1}{8\pi G_4 M}-\frac{2M A^2 G_4}{\pi}+{\cal O}(G_4^3M^3A^4)\,.
\end{eqnarray} 
The leading term is  exactly the contribution from the static black hole \cite{Emparan:1999wa}, while the second term is a finite temperature correction.

\begin{figure}[ht]
\leftline{
\includegraphics[width=0.5\textwidth]{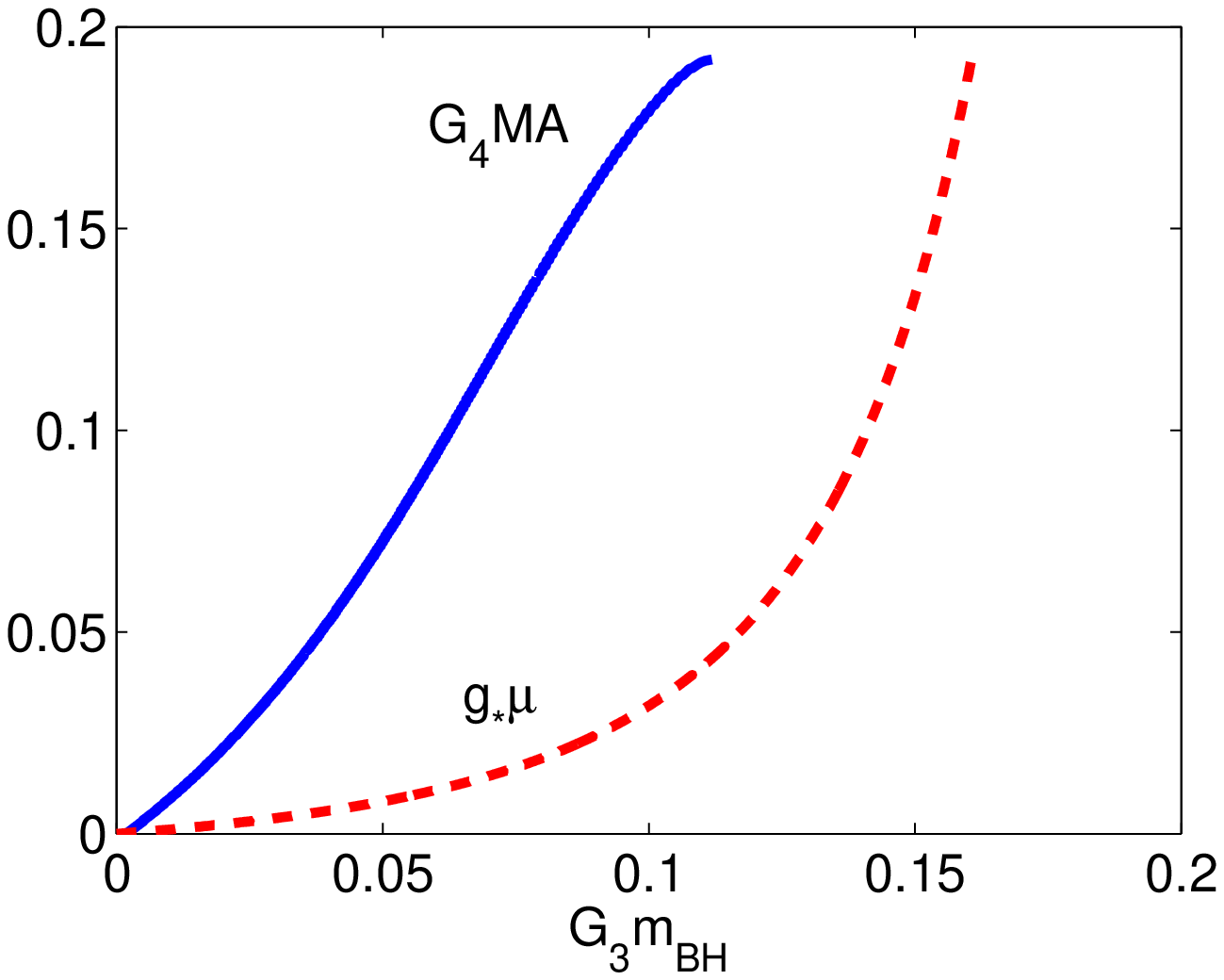}
\includegraphics[width=0.5\textwidth]{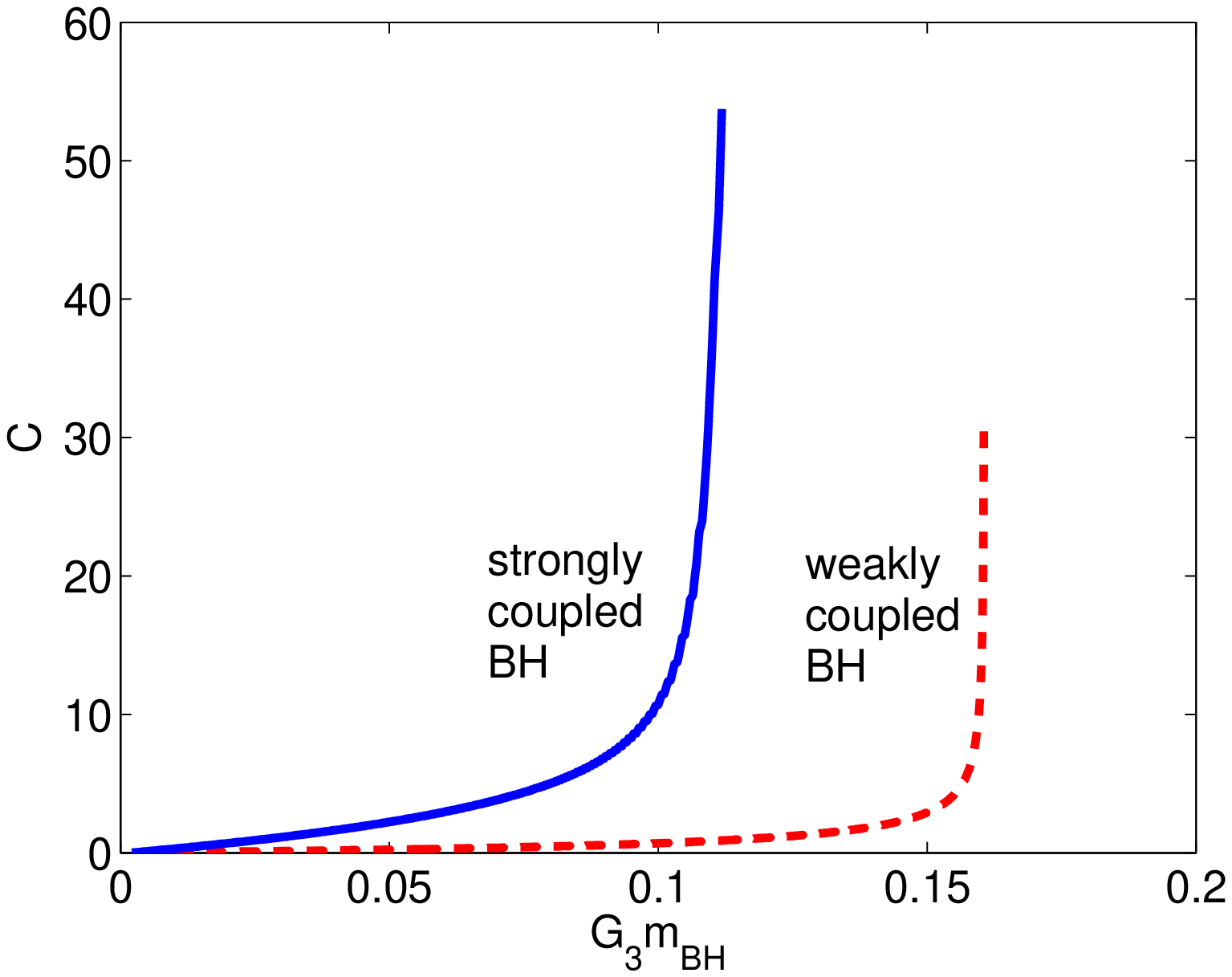}
}
\leftline{
\includegraphics[width=0.5\textwidth]{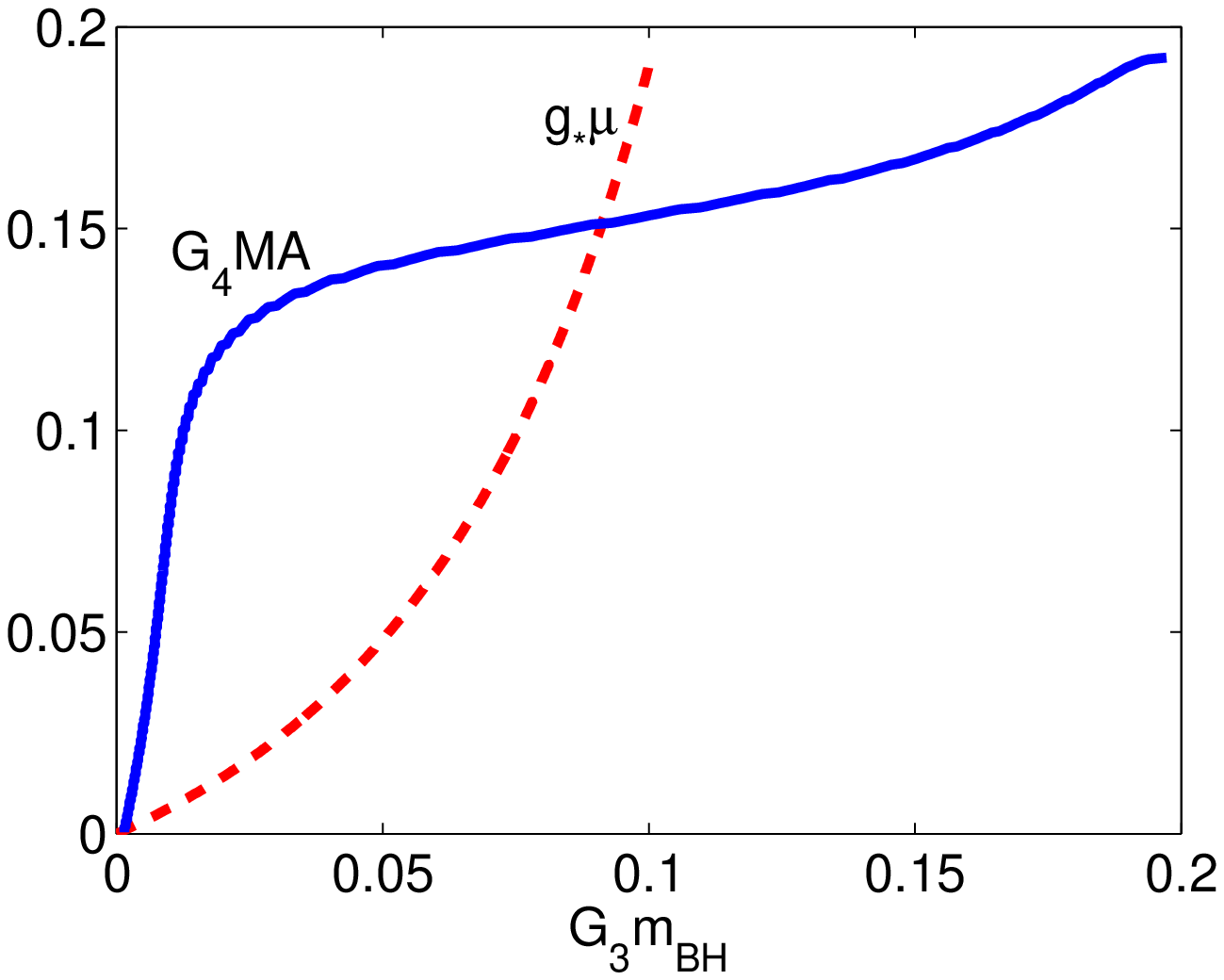}
\includegraphics[width=0.5\textwidth]{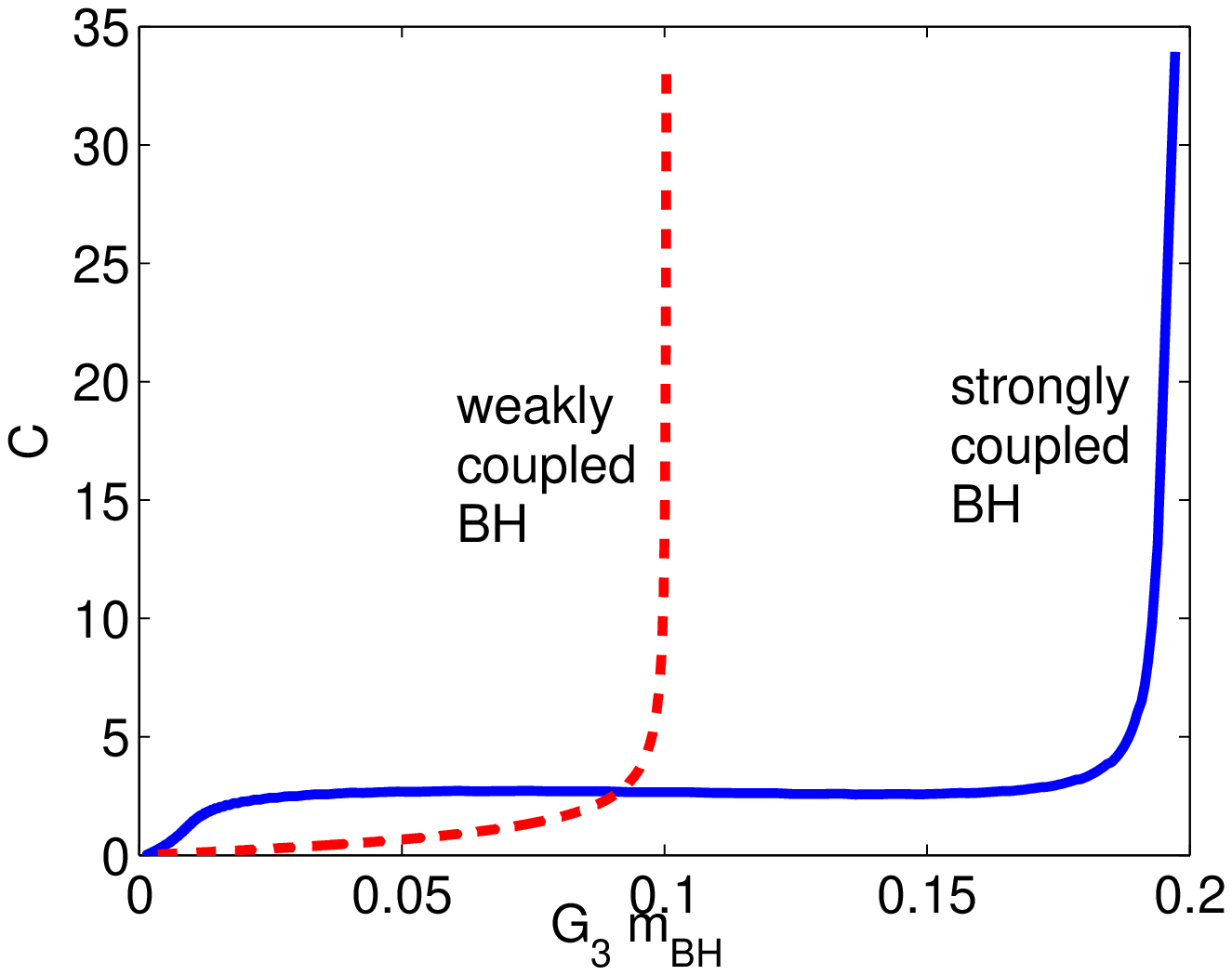}
}

\caption{On the left, the weakly and strongly coupled CFT parameters $g_{*}\mu$ and $G_4MA$ respectively as functions of the black hole mass. On the right, the corresponding proper circumference of the event horizon for both SCFTBH and WCFTBH. We take $\lambda=-0.2$, $g_{*}\approx 16$, and $\lambda=-0.9$, $g_{*}\approx 125$ for the upper and lower graphs, respectively. }
\label{Strong and weak CFT}
\end{figure}

According to AdS/CFT, the duality in the case of a 2-brane is between M-theory on $AdS_4 \times S^7$ and the CFT theory describing the world volume dynamics of a large number, $N$, of $M_2$ branes \cite{Petersen:1999zh}. In this case the effective number of degrees of freedom of the CFT  is given by  $g_{*}\sim N^{3/2}\sim \ell_4/ G_3$. Using the relations $G_3=G_4/2\ell_4$ and $\ell_4=1/\tilde A\sqrt{1+\lambda}$ we obtain
\begin{equation}\label{NDOF CFT}
g_{*}\sim\frac{2}{G_4\tilde A^2 (\lambda+1)}\,.
\end{equation}
 $g_{*}$ is taken to be a large number in order to suppress the quantum corrections to the supergravity approximation of M-theory, which results in a strongly coupled theory on the CFT side. Further, to make a connection between the strongly and weakly coupled solutions one has to consider the same number of degrees of freedom in both theories. This can be done by taking $g_{*}$ times  the energy-momentum tensor of the weakly coupled CFT in (\ref{the final answer of the energy-momentum tensor}).

\begin{figure}[ht]
\centerline{
\includegraphics[width=0.5\textwidth]{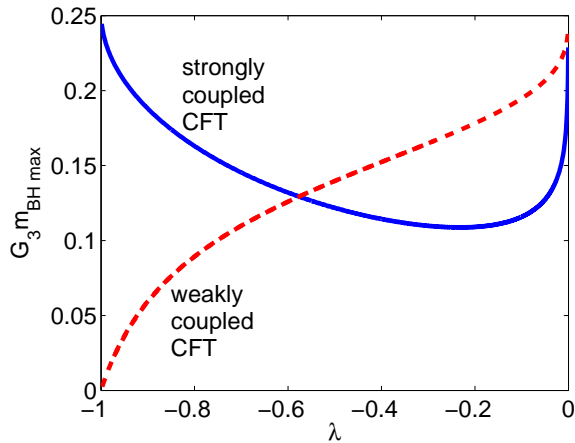}
}
\caption{The maximum mass a black hole can attain versus $\lambda$ for both SCFTBH and WCFTBH. The region under each curve represents the allowed values of the black hole mass.}
\label{MBHMAX V LAMBDA}
\end{figure}

 Now, we can use eqs. (\ref{equation of the brane}), (\ref{definition of theta}) and (\ref{NDOF CFT}) as well as numerical techniques to show that the functional dependence of $G_4MA$ and $g_{*}\mu$ is given by
\begin{eqnarray}
\nonumber
G_4MA&=&G_4MA(\lambda,G_3m_{\mbox{\scriptsize BH}})\,,\\
g_{*}\mu&=&g_{*}\mu(\lambda, G_3m_{\mbox{\scriptsize BH}})\,.
\end{eqnarray}
 In Figure (\ref{Strong and weak CFT}), we plot the strongly and weakly coupled CFT parameters $G_4MA$ and $g_{*}\mu$ respectively versus the black hole mass for small and large values of $\lambda$, or equivalently for low and high background temperatures reminding that $T=\sqrt{-\lambda}\tilde A/2\pi$. In both cases $G_4MA$ and $g_{*}\mu$ are increasing functions of $m_{\mbox{\scriptsize BH}}$. However, for  small $\lambda$ we find that SCFTBH space closes up (the two zeros $w_1$ and $w_2$ coincides) at a smaller value of $m_{\mbox{\scriptsize BH}}$ compared to that of WCFTBH. This happens when $G_4MA$ or $g_{*}\mu $ approaches the critical value $1/3\sqrt{3}$. 
In the same figure, we see  that for small values of $\lambda$ SCFTBH has larger proper horizon circumference  than WCFTBH. This matches exactly what one would find in the case of the $2+1$ D static black hole constructed previously in \cite{Emparan:1999wa} by cutting a AdS C-bulk having $\lambda=0$ with a critical brane. Once again, as one approaches the critical number $1/3\sqrt{3}$, the two zeros $w_1$ and $w_2$  converge to the same value which results in divergent horizon circumference. For $\lambda \approx -1$, the  weakly and strongly coupled solutions exchange their behavior to find that the space of the former closes up and hence its horizon circumference diverges at smaller values of $m_{\mbox{\scriptsize BH}}$. The transition from small to large $\lambda$ regime happens at $\lambda \approx -0.6$ ($T \approx 0.12\tilde A$) as can be seen in figure (\ref{MBHMAX V LAMBDA}) where we plot the maximum allowed black hole mass, before a naked singularity is formed, versus $\lambda$. The maximum possible mass for WCFTBH or SCFTBH  can approach $1/4G_3$ at small values of the background temperature. This result is expected since $1/4G_3$ is the maximum mass a static black hole can have in $2+1$ D. The maximum mass decreases monotonically with the temperature in the case of WCFTBH until a black hole ceases to exist when $T\rightarrow \tilde A/2\pi$. In contrast, there is a minimum of  $m_{\mbox{\scriptsize BH max}}\approx 0.1/G_3$ at $\lambda \approx -0.25$ ($T \approx 0.08 \tilde A$) in the case of SCFTBH. Beyond this temperature, the maximum mass increases  until it reaches $1/4G_3$ again at $T=\tilde A/2\pi$. This is a striking difference between  weakly and strongly coupled CFT.

\begin{figure}[ht]
\leftline{
\includegraphics[width=0.5\textwidth]{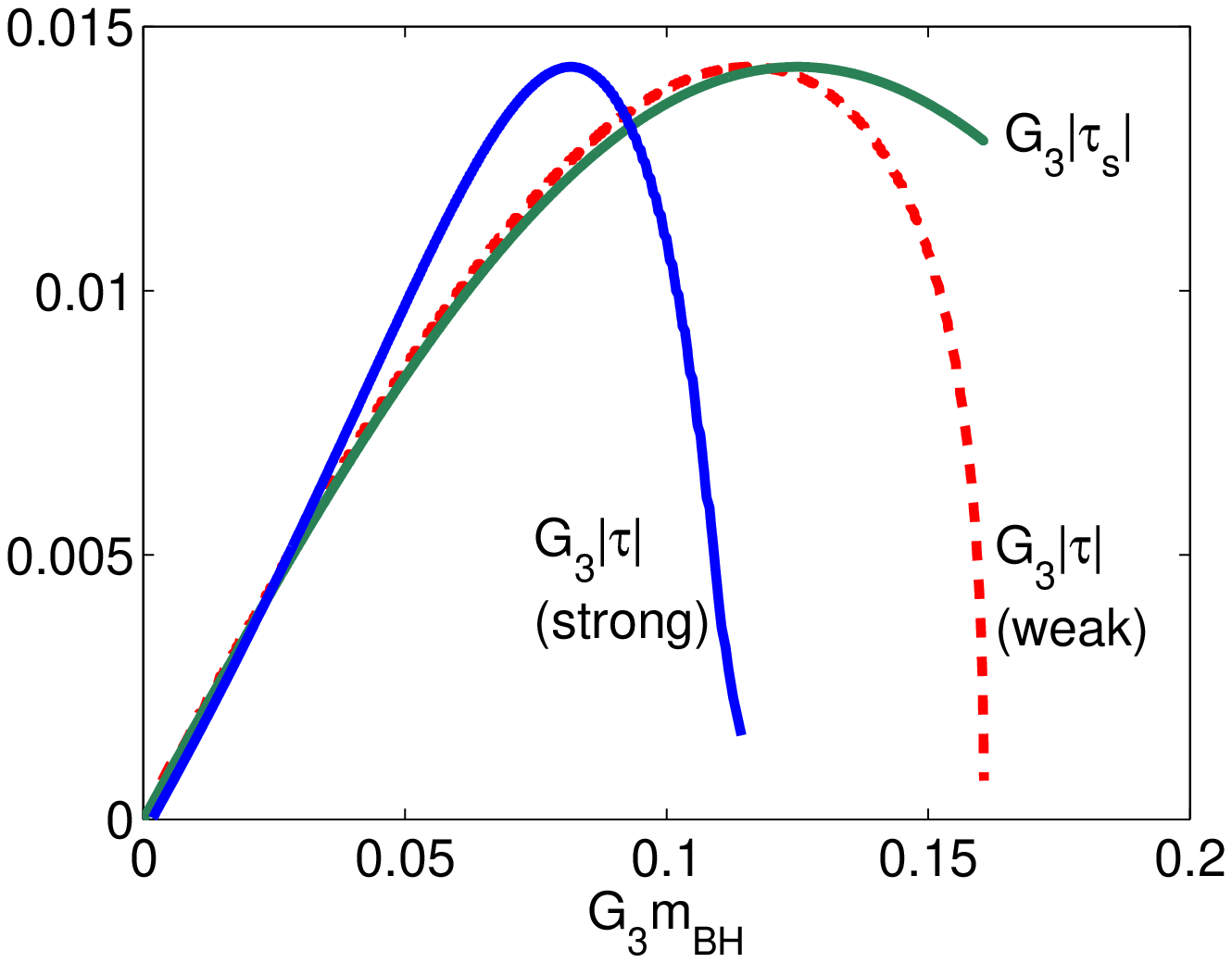}
\includegraphics[width=0.5\textwidth]{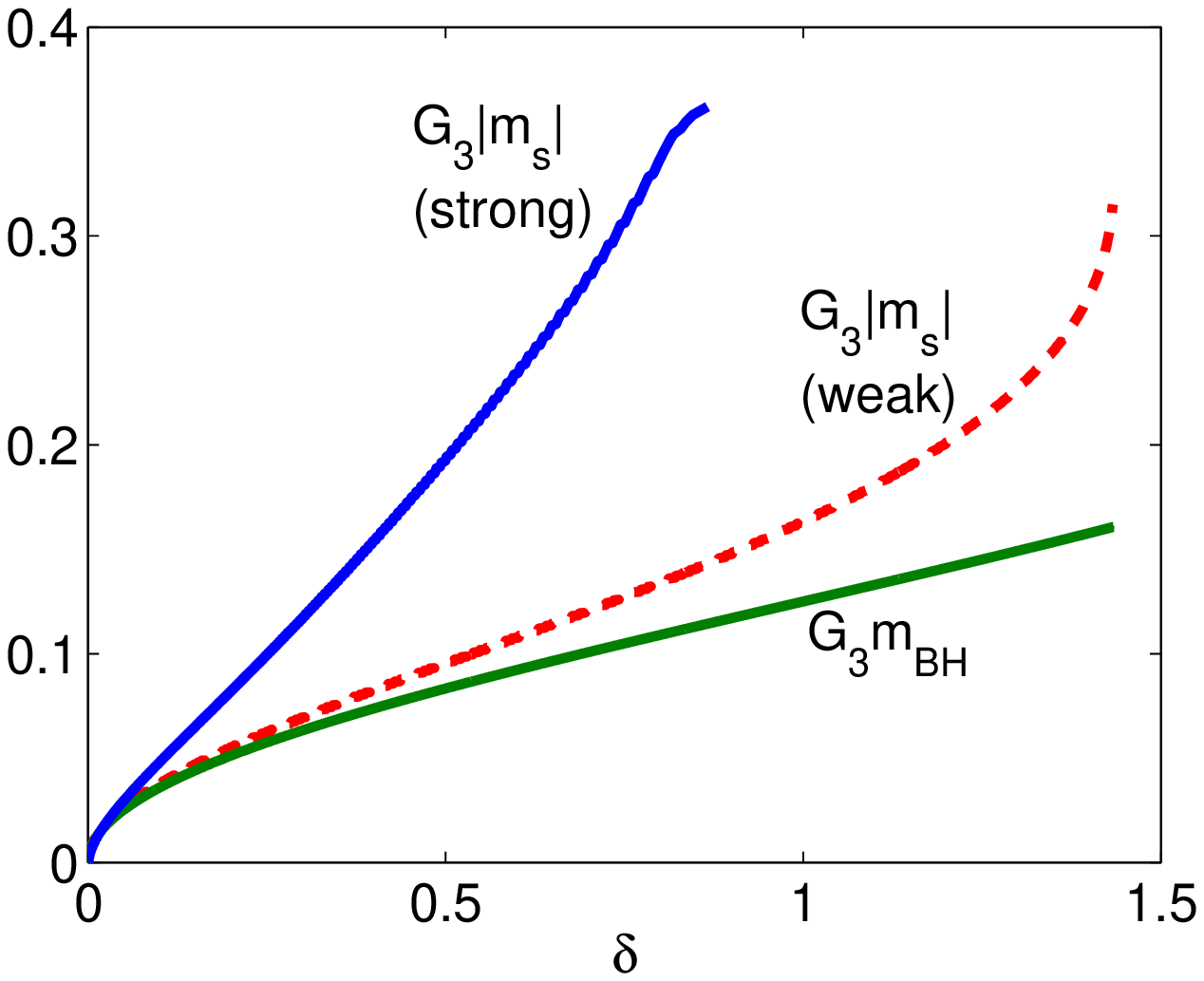}
}

\leftline{
\includegraphics[width=0.5\textwidth]{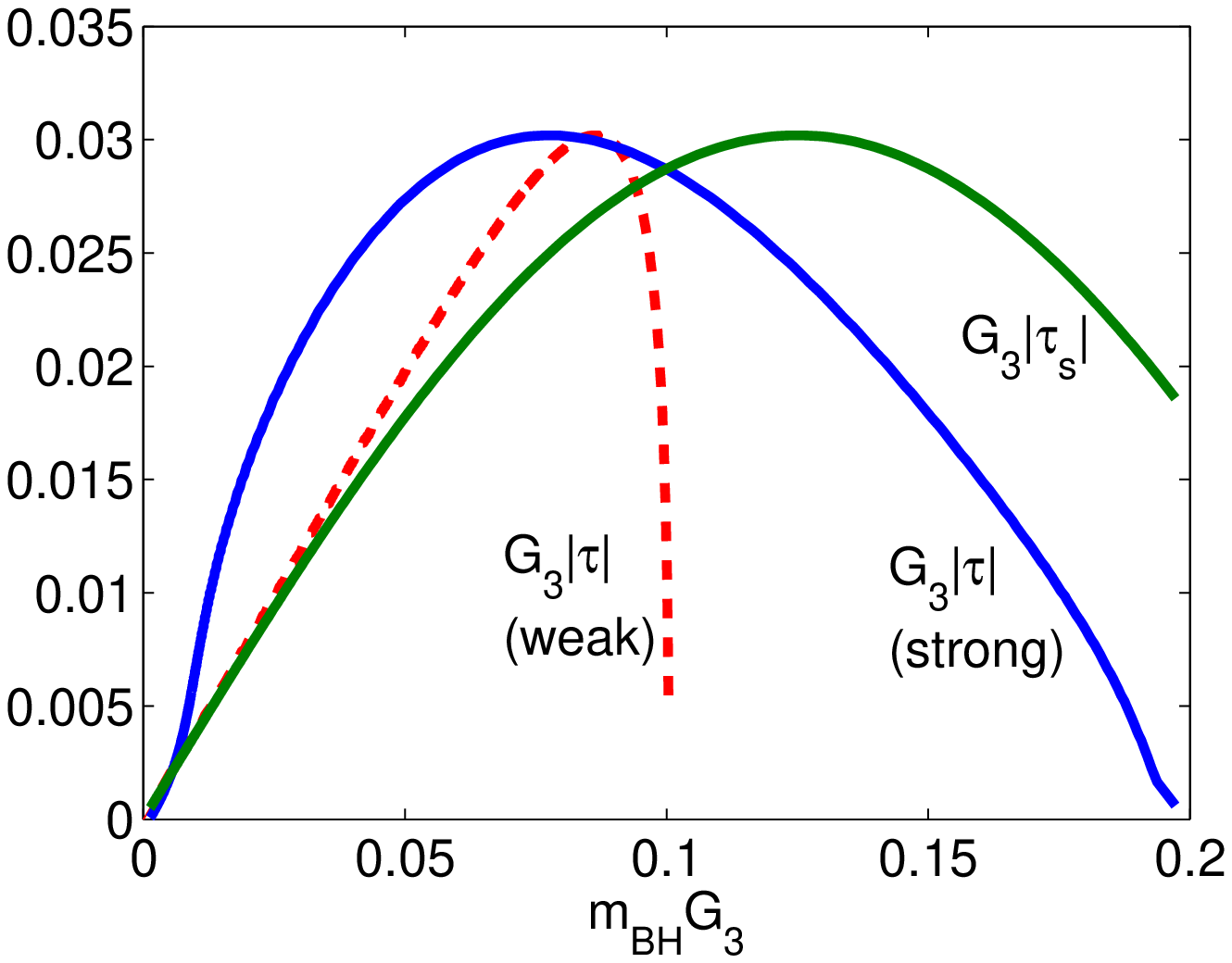}
\includegraphics[width=0.5\textwidth]{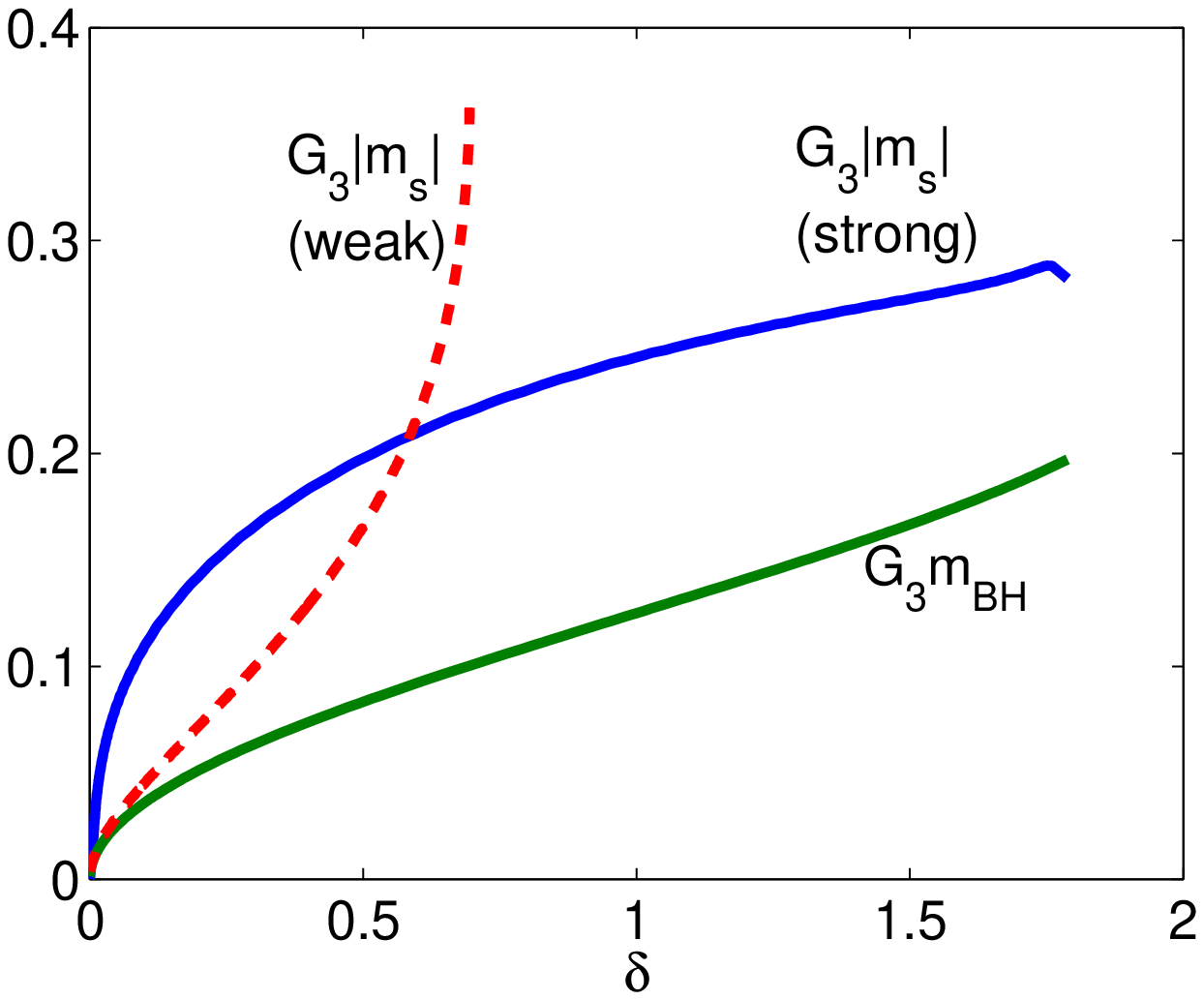}
}
\caption{On the left, we plot the classical value of the strut tension  $\tau_s$, given by (\ref{classical tension}), and the quantum corrected tension $\tau$ for weakly and strongly coupled CFT black holes. The quantum corrected value of the strut mass as well as the mass of the weakly and strongly coupled CFT black hole are plotted on the right. We take $\lambda=-0.2$, $g_{*}\approx 16$, and $\lambda=-0.9$, $g_{*}\approx 125$ for the upper and lower graphs, respectively.}
\label{Strong CFT tension}
\end{figure}

In figure (\ref{Strong CFT tension}), we plot the absolute value of the  classical tension $\tau_s$, and  quantum corrected tension $\tau$, given in (\ref{tension for strongly coupled CFT}), for both strongly and weakly coupled CFT black holes. The behavior of $\tau$ is qualitatively similar to $\tau_s$ as we find that the absolute value of the former increases with the black hole mass $m_{\mbox{\scriptsize BH}}$ until it reaches a maximum value of $A/4\pi G_3$. Further increase of $m_{\mbox{\scriptsize BH}}$ results in moving the position of the strut $w_4$ closer to $w_2$ and hence decreasing the value of $\tau$ until the space closes up when $w_2=w_1$. We also see that the absolute value of the quantum corrected tension is larger than the classical one for small values of the black hole mass, which is more evident in the case of strongly coupled CFT. This shows that the CFT is attractive for small values of $m_{\mbox{\scriptsize BH}}$: although gravity is not dynamical in $2+1$ D, the quantum corrections generate attractive force that tries to decelerate the black hole. Hence, one needs to increase the strut tension  to compensate for the attractive force. The picture is reversed for large values of $m_{\mbox{\scriptsize BH}}$ where we find that the CFT, instead, generates repulsive force. This can be understood by computing the mass of the strut as we show below.

 The mass of the quantum corrected strut reads $m_{s}=\tau \ell_{s}$, where  $\ell_s$ is the strut proper length  
\begin{equation}
\ell_s=\frac{1}{A}\int_{-\infty}^{w_1}\frac{dv}{(w_4-v)\sqrt{1-v^2-2G_4MAv^3}}+\frac{1}{A}\int_{w_1}^{w_2}\frac{dv}{(w_4-v)\sqrt{-1+v^2+2G_4MAv^3}}\,.
\end{equation} 
 From figure (\ref{Strong CFT tension}) , we see that contrary to the classical case where $|m_s|=m_{p}$, the quantum corrections renormalize the strut mass and hence violates this equality. Now, for small values of $m_{\mbox{\scriptsize BH}}$ we have $|m_s|\approx m_{p}$  and the CFT keeps its attractive nature. However, as we move to larger values of $m_{\mbox{\scriptsize BH}}$ we find  $|m_s|>m_{p}$, and bearing in mind the negative sign of $m_s$, the CFT reverses its nature which explains the repulsive force for large values of $m_{\mbox{\scriptsize BH}}$.

\section{Discussion and conclusion}

In this work, we have studied the CFT corrections to two accelerating masses moving with acceleration $A$, by means of  a strut connecting them, in a flat background. To achieve this, we first computed the thermal Green's function for a conformally coupled scalar. This function describes the quantum fluctuations of a quantum field kept in equilibrium with a thermal bath at temperature $T=A/2\pi$, which is the Hawking-Unruh temperature associated with the acceleration horizon. In calculating the Green's function, we used two different boundary conditions, namely, transparent and Dirichlet boundary conditions. We have found that the induced energy-momentum tensor diverges at the location of the strut in the case of Dirichlet boundary conditions, while the same quantity is regular for transparent boundary conditions. Further, we proceeded using the regular form of the energy-momentum tensor to calculate  the gravitational backreaction on the spacetime. Interestingly enough, we have shown that the  backreaction of the CFT  reproduces the same metric found before in \cite{Anber:2008qu} which was constructed  by cutting the $AdS_4$ spacetime with angular dependent critical brane. Although the former solution describes accelerated black hole dressed with weakly coupled CFT (WCFTBH), the latter, according to AdS/CFT+gravity, should describe a black hole dressed with strongly coupled CFT (SCFTBH). This is the first use of the duality in a system containing two interacting particles. Moreover, the presence of the CFT at finite temperature gave us an opportunity to study the finite temperature effects in a strongly coupled system. 

The existence of the brane implies that high energy states in the dual CFT theory are integrated out, and hence the CFT has a UV cutoff scale $\mu_{\mbox{\scriptsize UV}}\sim 1/\ell_4$, or in terms of the  number of CFT degrees of freedom and the $3$ D Newton's constant $\mu_{\mbox{\scriptsize UV}}\sim 1/g_{*} G_3 $. The solution given by (\ref{black hole on brane strong CFT}) describes a genuine  black hole if the Compton's wavelength $\lambda_{C}=1/m_{\mbox{\scriptsize BH}}$ satisfies $\lambda_C < r_0$, where $r_{0}$ is the black hole horizon, otherwise quantum effects would smear it over a volume larger than the horizon radius. As was shown in \cite{Emparan:2002px}, this requirement sets a new scale $1/\sqrt{G_4}\sim 1/\sqrt{g_*}G_3$ on the brane. In other words, a black hole solution is reliable if its mass ranges in $\mu_{\mbox{\scriptsize UV}}<1/\sqrt{G_4}<m_{\mbox{\scriptsize BH}}<m_{\mbox{\scriptsize BH max}}$. The maximum black hole mass $m_{\mbox{\scriptsize BH max}}$ is equal to the maximum mass $1/4G_3$ one can place in $2+1$ D in the case of a static black hole. However, we found that $m_{\mbox{\scriptsize BH max}}$ is temperature dependent, and can be determined when the strongly or weakly coupled CFT parameters, $G_4MA$ and $g_{*}\mu$ respectively, reaches the critical value $1/3\sqrt{3}$. As was shown in figure (\ref{MBHMAX V LAMBDA}), this mass starts at $1/4G_3$ at low temperatures for both SCFTBH and WCFTBH, and then decreases monotonically until a black hole ceases to exist as $T \rightarrow\tilde A/2\pi$ in the case of WCFTBH. In contrast to this behavior, $m_{\mbox{\scriptsize BH max}}$ for SCFTBH develops a minimum  and then increases until it reaches $1/4G_3$ again at $T=\tilde A/ 2\pi$. This is a striking deference between strongly and weakly coupled CFT. Nevertheless, in both cases as the black hole mass reaches the maximum allowed value, the proper horizon circumference grows indefinitely. Beyond, $m_{\mbox{\scriptsize BH max}}$ the black hole horizon disappears leaving behind undressed singularity in a clear violation of the quantum cosmic censorship. In fact, it was shown in \cite{Anber:2008qu}  that the time-dependent solutions obtained there do not respect the quantum censorship, as well. Indeed, these results may indicate that the censorship operates only for a static CFT.

The presence of a naked singularity may signal phase transition on the bulk side. However, at this point, it is safe to say that a complete understanding of the nature of this singularity is still an open question. Further study of SCFTBH may reveal rich phenomena of finite temperature CFT coupled to gravity in $2+1$ D.

\section*{Acknowledgement}

The author is grateful to Lorenzo Sorbo for invaluable help and patience during this work, and to David Kastor for enlightening discussions. The author would like also to thank Yulia Smirnova for her careful reading of a previous version of the manuscript. This work has been supported by the U.S. National Science Foundation under the grant PHY-0555304.

%
\appendix
%

 \renewcommand{\theequation}{A\arabic{equation}}
  \setcounter{equation}{0}  
  \section*{Appendix A: Mass calculations}  

\begin{figure}[ht]
\leftline{
\includegraphics[width=0.5\textwidth]{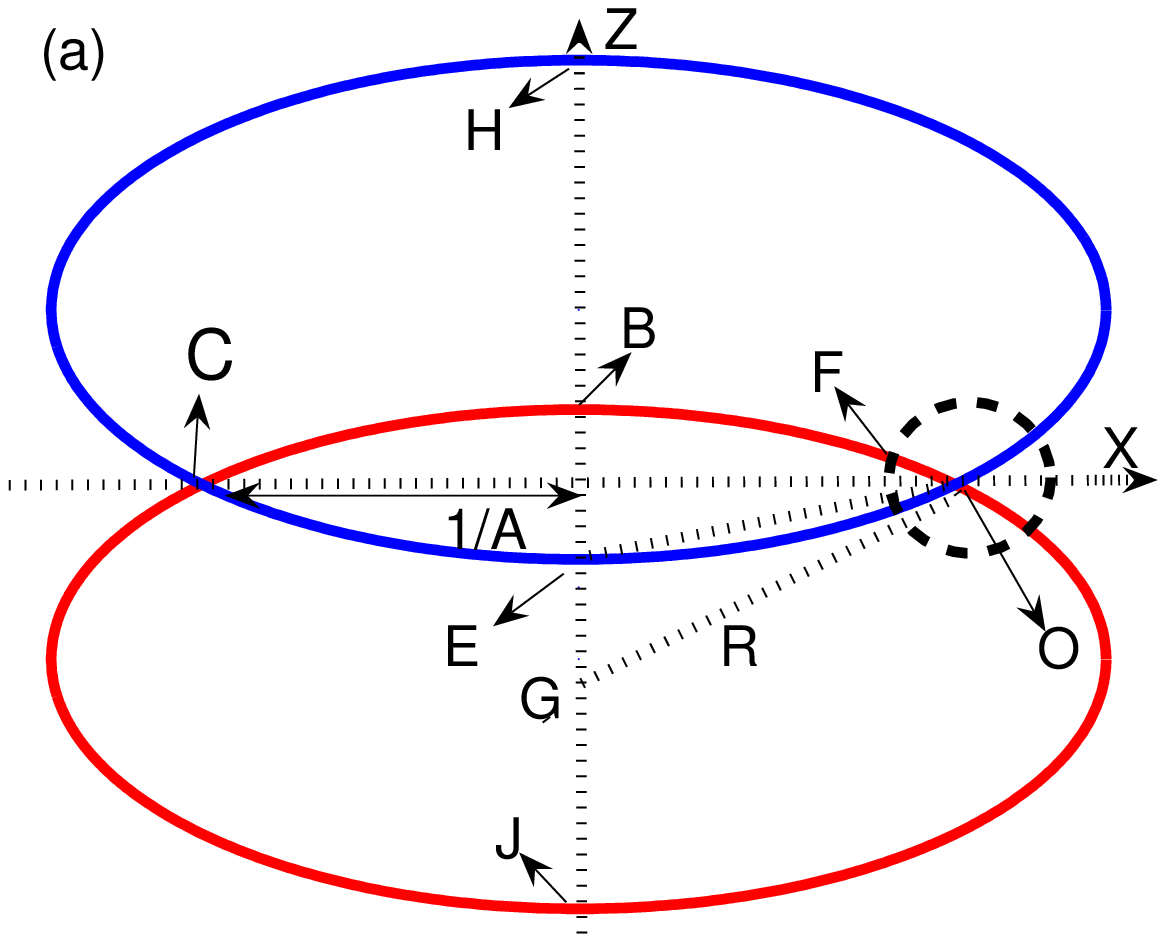}
\includegraphics[width=0.5\textwidth]{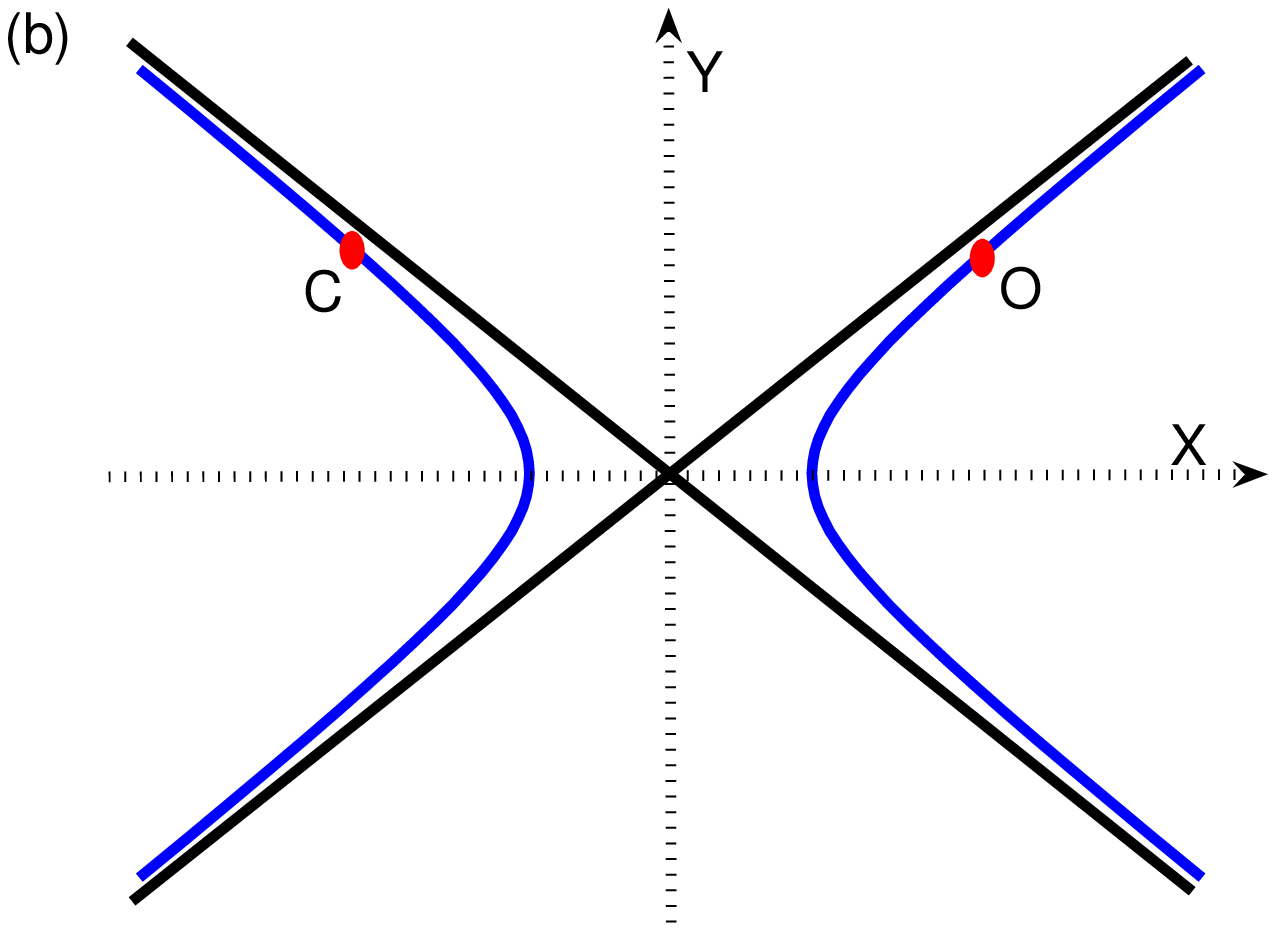}
}

\caption{In figure (a), we sketch the geometry in the $X-Z$ plane at time $Y=0$. In figure (b), the trajectories of the point particles $O$ and $C$ are shown in the $Y-X$ plane.}
\label{XYZ COORDINATES}
\end{figure}

In this appendix, we calculate the mass of the particles and strut as measured by a static observer sitting in Minkowskian background. We start from the coordinate transformations given in (\ref{transformations to flat metric}) which bring the metric in (\ref{accelerating conical sing}) to the form $ds^2=-dY^2+dX^2+dZ^2$. Further, we introduce the deficit parameter $\delta$ and require that $-\infty <t<\infty$, $v\le -1$ and $-1\le w\le 1-\delta$. These transformations cover only the portion of space  $\{X \ge 0\} \cup\{Z \ge 0\}$. To cover the whole space we take $w=-\cos\theta$ and $v=-\cosh\eta$, and demand that  $-\infty<\eta<\infty$, and $-\pi/p\le \theta \le \pi/p$, where the parameter $p$ is defined as $p=\pi/\cos^{-1}(-1+\delta)$. Now, taking the limit $\eta \rightarrow \pm \infty$ we obtain
\begin{eqnarray}
\nonumber
X=\pm\frac{\cosh t}{A}\,,\\
\nonumber
Y=\pm\frac{\sinh t}{A}\,,\\
Z=0\,.
\end{eqnarray}
This describes two point particles moving along the hyperbola $X^2-Y^2=1/A^2$, as shown in figure (\ref{XYZ COORDINATES}). In other words it describes two point particles accelerating in the plane $Z=0$, with acceleration $A$, in two opposite directions. In order to complete the picture, we need to understand the role of the deficit parameter $\delta$ in the $XYZ$ coordinates. This can be achieved by eliminating $v$ and $t$ using the set of transformations given in (\ref{accelerating conical sing}) to find
\begin{equation}\label{large circles equation}
X^2+\left(Z\mp \frac{w}{A\sqrt{1-w^2}}\right)^2=Y^2+\frac{1}{A^2(1-w^2)}\,,
\end{equation}
which describe two circles, in the $XZ$ plane, with origin located at 
$G=(0,\pm w/A\sqrt{1-w^2})$ and radius ${\cal R}^2=Y^2+1/A^2(1-w^2)$. A sketch of the loci of the circles is provided in figure (\ref{XYZ COORDINATES}).  For small values of $\delta$, the intersection region $OBCEO$ is small, and  the final space is constructed by removing this region and gluing the two curves $OBC$ and $OEC$ taking $w=w_0=1-\delta$. Hence, the strut is  interpreted as the curve $OBC$ or equivalently the curve $OEC$. This strut stretches between the points $O$ and $C$ where the two circles intersect. These points move along the hyperbola described above and hence we can assign point particles at the location of these points. As we tune $\delta$ to higher values, the size of the intersection region increases until $\delta=1$. At this value of $\delta$, the two circles coincide. For $\delta>1$, the blue and the red circles in figure (\ref{XYZ COORDINATES}) exchange their positions, and the space in this situation is constructed by removing the region $OHCBO\cup OJCEO$, and then gluing the curves $OHC$ and $OJC$, until the space disappears when $\delta=2$. 
In summary, we have two point particles located at $A$ and $O$ which are accelerated by means of a strut stretched between them and pushing them away. Since mass in $2+1$ D is interpreted as missing space, this configuration is realized by removing the region $OBCEO$ for $\delta<1$ or $OHCBO\cup OJCEO$ for $1<\delta<2$  shown in figure (\ref{XYZ COORDINATES}). In the following, we perform the calculations assuming $\delta<1$. However, the results are valid even for $1<\delta<2$.

The length of the arc $OB$ (or equivalently $OE$) at time $Y=0$ can be calculated easily from the geometry of figure (\ref{XYZ COORDINATES}) to find 
\begin{equation}
\mbox{Arc length}=\frac{\pi-\cos^{-1}(-w_0)}{A\sqrt{1-w_0^2}}\,.
\end{equation}
Moreover, we have shown in section 2 that the strut tension, which is a Lorentz invariant, is given by $\tau_s=-A\sqrt{2\delta-\delta^2}/4\pi G_3$. Hence, we find that the strut mass, $m_s= \tau_s\times \mbox{Arc length}$, reads
\begin{equation}
m_s=-\frac{1}{4G_3}\left[1-\frac{\cos^{-1}\left(-1+\delta\right)}{\pi} \right]\,,
\end{equation}
which is exactly what we have found in section 2 working in the $vw$ coordinates. From  the point of view of an observer in the $XYZ$ coordinates, the length of the strut and hence its mass will increase with time. However, an observer moving with one of the point particles, at $O$ or $C$, will conclude that the mass of the strut is constant according to her sticks and clocks.

On the other hand, the mass of the point particles located at $O$ or $C$ can be determined, according to our $XYZ$ observer, by measuring the circumference of a circle with radius $r$ centered at $O$ or $C$, and further dividing the result by $2\pi r$ to get the total angle enclosed $\Delta \theta$. Then, the mass is given by $m_{p}=(2\pi-\Delta\theta)/8\pi G_3$. To this end, consider a circle of radius $r$ centered at point $O$ at time $Y=0$
\begin{equation}\label{smaller circle}
\left(X-\frac{1}{A}\right)^2+Z^2=r^2\,.
\end{equation}   
This circle intersects (\ref{large circles equation}) in $F=\left(1/A+\sqrt{r^2-Z_0^2},Z_0\right)$ where
\begin{equation}
Z_0=\frac{2aA^2r^2+r\sqrt{(4a^2-r^2)A^2+4}}{2(a^2A^2+1)}\,,
\end{equation}
and $a=w_0/A\sqrt{1-w_0^2}$. Using simple geometric calculations we find that the angle between the positive $X$-axis and the location of $F$ is given by
\begin{equation}
\alpha=\tan^{-1}\left(\frac{Z_0}{\sqrt{r^2-Z_0^2}}\right)\,.
\end{equation}
In the limit $r\rightarrow 0$ we obtain
\begin{equation}
\alpha=\mbox{lim}_{r\rightarrow 0}\tan^{-1}\left(\frac{Z_0}{\sqrt{r^2-Z_0^2}}\right)=\tan^{-1}\left[\frac{\sqrt{1-w_0^2}}{w_0}\right]=\cos^{-1}\left(-1+\delta\right)\,,
\end{equation}
and hence the total angle enclosed by the circle is $\Delta\theta=2\alpha$, and the particle mass is given by
\begin{equation}
m_p=\frac{1}{4G_3}\left[1-\frac{\cos^{-1}\left(-1+\delta\right)}{\pi} \right]\,.
\end{equation}
Hence, we find $m_s+m_p=0$ as was shown in section 2. This can be justified by enclosing the geometrical construction by very large circle at infinity to find that the total deficit angle is actually zero.

 \renewcommand{\theequation}{B\arabic{equation}}
  \setcounter{equation}{0}  
  \section*{Appendix B: Properties of associated Legendre functions }  
In this appendix, we summarize the main properties and relations of associated Legendre's functions used throughout the paper \cite{bateman,morse,Snow}. The associated Legendre equation of degree $\nu$ and order $\mu$ reads
\begin{equation}
(1-v^2)\frac{d^2 G}{d v^2}-2v\frac{d G}{d v}+\left[\nu(\nu+1)-\frac{\mu^2}{1-v^2}\right]G=0\,,
\end{equation} 
where we assume $v\ge 1$. The solutions of this equation are given by  $G= P^{\pm\mu}_{\nu}(v)$ and $Q^{\pm\mu}_{\nu}(v)$, the associated Legendre's functions of the first and second kind respectively. For integer values of $\mu$, we find that the only well-behaved functions at $v=1$ and $v=\infty$ are $P^{-m}_{\nu}(v)$ and $Q^{-m}_{\nu}(v)$ respectively, where $m=0,1,2,...$. The Wronskian of these functions is given by 
\begin{eqnarray}
W\{P^{-m}_{\nu},Q^{-m}_{\nu} \}&=&
P^{-m}_{\nu}\frac{d Q^{-m}_{\nu}}{dv}-Q^{-m}_{\nu}\frac{d P^{-m}_{\nu}}{dv}\\
&=&\frac{(-1)^m}{(1-v^2)}\frac{\Gamma(1-m+\nu)}{\Gamma(1+m+\nu)}\,.
\label{Wronskin}
\end{eqnarray}
The associated Legendre's functions satisfy the addition theorem
\begin{eqnarray}
\nonumber
&Q_{\nu}&\left[vv'-\sqrt{v^2-1}\sqrt{v'^2-1}\cos(\phi-\phi') \right]=\\
&\sum_{m=0}^{\infty}&\epsilon_{m}(-1)^m\frac{\Gamma(1+m+\nu)}{\Gamma(1-m+\nu)}P^{-m}_{\nu}(v_<)Q^{-m}_{\nu}(v_>)\cos m(\phi-\phi')\,.
\label{addition theorem}
\end{eqnarray}
To sum up the series (\ref{main series}) we use the integral representation
\begin{equation}\label{integral representation of Q}
Q_{\nu}(\cosh\gamma)=\frac{1}{\sqrt{2}}\int_{\gamma}^{\infty}\frac{e^{-(\nu+1/2)u}}{\sqrt{\cosh u-\cosh \gamma}}.
\end{equation}
Finally, for $\nu=m-1/2$ one can make use of the identity
\begin{equation}\label{identity for Legendre functions}
\frac{1}{\sqrt{z-\cos(\theta\pm\theta')}}=\frac{\sqrt{2}}{\pi}\sum_{m=0}^{\infty}\epsilon_{n}Q_{m-1/2}(z)\cos m(\theta\pm\theta')\,.
\end{equation}
%

\renewcommand{\theequation}{C\arabic{equation}}
  \setcounter{equation}{0}  
  \section*{Appendix C: Calculations of the energy-momentum tensor }  


The energy-momentum tensor is given by the coincidence limit (\ref{coincidence limit}). The coincidence limits of the various derivatives in the case of transparent boundary conditions are given by
\begin{eqnarray}
\nonumber
\frac{4\sqrt{2}\pi^2}{A}\mbox{lim}_{x'\rightarrow x}\frac{\partial G}{\partial \eta}&=&\frac{I_1(p)}{2}\sinh\eta \\
\nonumber
\frac{4\sqrt{2}\pi^2}{A}\mbox{lim}_{x'\rightarrow x}\frac{\partial G}{\partial \theta}&=&\frac{ I_1(p)}{2}\sin\theta\\
\nonumber
\frac{4\sqrt{2}\pi^2}{A}\mbox{lim}_{x'\rightarrow x}\frac{\partial^2 G}{\partial \eta\partial\eta'}&=&\frac{ I_1(p)\sinh^2\eta}{4(\cosh\eta-\cos\theta)}+I_3(p)(\cosh\eta-\cos\theta)\\
\nonumber
\frac{4\sqrt{2}\pi^2}{A}\mbox{lim}_{x'\rightarrow x}\frac{\partial^2 G}{\partial \theta\partial\theta'}&=&\frac{I_1(p)\sin^2\theta}{4(\cosh\eta-\cos\theta)}-I_2(p)(\cosh\eta-\cos\theta) \\
\nonumber
\frac{4\sqrt{2}\pi^2}{A}\mbox{lim}_{x'\rightarrow x}\frac{\partial^2 G}{\partial \phi\partial\phi'}&=&I_3(p)\sinh^2\eta\left(\cosh\eta-\cos\theta\right)\\
\nonumber
\frac{4\sqrt{2}\pi^2}{A}\mbox{lim}_{x'\rightarrow x}\frac{\partial^2 G}{\partial \eta^2}&=&I_1(p)\left (\frac{\cosh\eta}{2}-\frac{\sinh^2\eta}{4(\cosh\eta-\cos\theta)} \right)-I_3(p)\left(\cosh\eta-\cos\theta\right)\\
\nonumber
\frac{4\sqrt{2}\pi^2}{A}\mbox{lim}_{x'\rightarrow x}\frac{\partial^2 G}{\partial \theta^2}&=&I_1(p)\left (\frac{\cos\theta}{2}-\frac{\sin^2\theta}{4(\cosh\eta-\cos\theta)} \right)-I_2(p)\left(\cosh\eta-\cos\theta\right)\,,\\ 
\label{coincidence limits of derivatives}
\end{eqnarray} 
where 
\begin{eqnarray}
\nonumber
 I_{1}(p)&=&\sqrt{2}\int_{0}^{\infty}\frac{du}{\sinh u}\left[p\,\coth p\,u -\coth u \right]\\
\nonumber
 I_{2}(p)&=&-\frac{1}{\sqrt{2}}\int_{0}^{\infty}\frac{du}{\sinh u}\left[p^3\frac{\cosh p\,u}{\sinh ^3 p\,u}-\frac{\cosh u}{\sinh ^3 u}  \right]\\
\nonumber
 I_{3}(p)&=&\int_{0}^{\infty}\frac{du}{\sqrt{\cosh u-1}}\left[\frac{p^2}{2\sinh u \sinh^2(p\,u/2)}+\frac{p\cosh u\coth(p\,u/2)}{\sinh^2 u}-(p\leftrightarrow 1) \right]\,.\\
\label{integrals per}
\end{eqnarray}
 Moreover, using the fact that the two point function satisfies the homogeneous equation $g_E^{\mu\,\nu}\nabla_{\mu}\nabla_{\nu}G^{R}(x,x')=0$ outside the singularity we get the consistency condition  $ I_3(p)=I_2(p)/2+I_1(p)/8$. Now, substituting the relations (\ref{coincidence limits of derivatives}) in eq. (\ref{coincidence limit}) we obtain the energy-momentum tensor 
\begin{equation}
\left\langle T_{\nu\,\mbox{\scriptsize TR}}^{\mu}(x) \right\rangle=\frac{A^3 I_{2}(p)}{8\sqrt{2}\,\pi^2}\left(\cosh\eta-\cos\theta \right)^3\mbox{diag}(1,1,-2)\,.
\end{equation}

One can follow the same procedure above to obtain in the case of Dirichlet boundary conditions
\begin{equation}\label{ energy-momentum tensor Dirchilet}
\left\langle T_{\nu\,\mbox{\scriptsize D}}^{\mu}(x) \right\rangle=\frac{A^3 I(\theta,p)}{16\sqrt{2}\pi^2}\left(\cosh\eta-\cos\theta \right)^3\mbox{diag}(1,1,-2)\,,
\end{equation}
where
\begin{eqnarray}
\nonumber
I(\theta,p)&=&\int_{0}^{\infty}du\frac{p^3\sinh (p\,u/2)\cos p\,\theta }{4\sqrt{\cosh u-1}(\cos 2\,p\,\theta-\cosh p\,u)^3}\times\\
\nonumber
&\times&\left(-14+8\,\cos 2\,p\,\theta+\cos 4\,p\,\theta+12\cos 2\,p\,\theta \cosh p\,u-8\cosh p\,u+\cosh 2\,p\,u\right)\\
&-&\frac{1}{\sqrt{2}}\int_{0}^{\infty}\frac{du}{\sinh u}\left[\frac{p^3}{\sinh^3 p\,u}+\frac{p^3}{2\,\sinh p\,u} \right] -(p\rightarrow 1)\,.
\label{integral D}
\end{eqnarray}

\end{document}